\documentclass[preprint]{aastex}

\shorttitle{Pinsonneault et al.}
\shortauthors{Main-sequence fitting}

\begin{document}

\title{The Distances to Open Clusters as Derived from
Main-sequence Fitting. II.  Construction of Empirically
Calibrated Isochrones\footnote{This publication makes
use of data products from the Two Micron All Sky
Survey, which is a joint project of the University of
Massachusetts and the Infrared Processing and Analysis
Center/California Institute of Technology, funded by the
National Aeronautics and Space Administration and the
National Science Foundation.} }

\author{Marc H. Pinsonneault and Donald
M. Terndrup} \affil{Department of Astronomy,
The Ohio State University, Columbus, OH
43210}\email{pinsono,terndrup@astronomy.ohio-state.edu}

\author{Robert B. Hanson} \affil{University of
California Observatories/Lick Observatory, Santa Cruz,
CA 95064}\email{hanson@ucolick.org}

\and

\author{John R. Stauffer} \affil{Infrared Processing and
Analysis Center, California Institute of Technology, Mail
Code 100-22, 770 South Wilson Avenue, Pasadena, CA 91125}
\email{stauffer@ipac.caltech.edu}

\begin{abstract} We continue our series of papers on open
cluster distances by comparing multicolor photometry of
single stars in the Hyades to theoretical isochrones
constructed with various color-temperature relations.
After verifying that the isochrone effective temperatures
agree well with spectroscopically determined values,
we argue that mismatches between the photometry and the
theoretical colors likely arise from systematic errors
in the color-temperature relations.  We then describe a
method for empirically correcting the isochrones to match
the photometry, and discuss the dependence of the isochrone
luminosity on metallicity.  \end{abstract}



\section{Introduction}

Open clusters are important laboratories for testing
stellar evolution models and for deciphering the star
formation history of the Galaxy, since each cluster
contains samples of stars of a single age and (probably)
composition.  As is well known, determining distances is
the most fundamental step to measuring cluster ages and
other key properties.

The success of the Hipparcos mission \citep{esa97}
has now provided us with many cluster distances derived
from stellar parallaxes.  While the new astrometry is
a significant advance, it did not come without some
controversy.  In particular, there were a few cases,
notably the Pleiades, in which the Hipparcos parallax
distance \citep{mer97,rob99,vl99} was significantly
different from that obtained by main-sequence (MS) fitting
using stellar isochrones \citep{pin98,sod98}.  A number
of papers have discussed these discrepancies in cluster
distance estimates, which could have several causes.
First, the systematic or random errors in MS-fitting
distances could be significantly larger than estimated,
as claimed by \citet{rob00}, or the photometry, reddening,
and/or metal abundances in the best-studied clusters
could be seriously in error.  In the case of the Pleiades,
the cluster may have an unusually high helium abundance,
making its main sequence $\approx 0.3$ mag fainter than
would be expected from its metallicity \citep{bel98}; this,
however, is difficult to understand since there do not seem
to be nearby field stars of similar characteristics in the
Hipparcos catalog \citep{sod98}.  Alternatively, the metal
abundance from spectroscopy may have been overestimated
\citep[][and references therein]{per03};  this point has
been disputed by, {\it inter alia}, \citet{sn01}.  Second,
there could be some overlooked flaw in the stellar models
used to compute isochrones or in the transformation of
model quantities to broadband colors.  Particularly on
the lower main-sequence in young clusters, stellar
activity producing spots or chromospheric emission can
affect colors in the blue \citep{vam87,sta03}.  Finally,
the cluster parallax discrepancies, amounting to $\sim
1$ mas, may just reflect the actual size of Hipparcos'
parallax errors over the small angular scale ($\lesssim
1^\circ$) subtended by most open clusters except the
Hyades.  On scales comparable to Hipparcos' $0{\fdg}9$
field of view, the individual stellar parallaxes are
correlated \citep{le98}, causing local zero-point errors
in the cluster parallaxes.  These have been suggested
to be $\sim 1$ mas in the Pleiades \citep{makarov02}
and Coma Ber (Makarov 2003, preprint), but smaller in
the Hyades \citep{ng99a,ng99b,bhz01}.  Recently, there
has been a similar discussion about the distance to
NGC~2516 \citep{terndrup02,per03};  here, uncertainties
in the cluster metallicity and in the amount of foreground
extinction complicate the comparison between the MS-fitting
distance and the Hipparcos measurement.  Regardless of the
source or sources of these discrepancies, it remains the
case that the 1997-1999 analysis of Hipparcos parallaxes
and current stellar evolutionary models are not mutually
consistent when applied to the ensemble of nearby clusters
with precise parallaxes.

The idea that MS-fitting can be used to find distances
to clusters and even individual stars has, of course,
been around for a long time \citep[e.g.,][]{eggen48,jk55},
but it has proven difficult to develop a precise technique
using theoretical isochrones, even for determining relative 
distances\footnote{Alternative methods using field
stars with measured parallaxes have equivalent challenges;
see, for example, \citet{dvb00}.}.  The stellar
evolution models themselves are now well constrained
in the case of the Sun \citep[e.g.,][]{bpb00,bpb01},
but are less certain for the full range of temperatures
and luminosities in any open cluster.  In order to
generate isochrones, the models need transformations
from theoretical quantities ($L, T_{\rm eff}$) into
observational magnitudes and colors.  A number of different
color-temperature calibrations are available, including
some generated empirically \citep[e.g.,][]{gr96,ws99,ls01,vc03}.
Despite improvements in the determination of atmospheric
opacities, bolometric corrections, etc., there typically
remain small but significant mismatches between calculated
and isochrone colors and the best photometry in open
clusters \citep[e.g.,][]{ter00,bhz01,cas01,gs03}.  The size
of these discrepancies can be as large as 0.04 $-$ 0.06
mag in broadband colors.  Furthermore, the dependence on
metallicity of the absolute magnitude $M_V$ of the MS is a
function of color, being greater when bluer colors such as
$B - V$ are used in the color-magnitude diagram (CMD) and
less, for example, when $V - I$ or $V - K$ are employed.
Taken together, all this means that the derived distance
depends on the selection of colors used in the MS-fitting
\citep{pin98}.  On the other hand, if the isochrones can
be accurately calibrated over a wide range of colors, it
may be possible to use multiband photometry to derive more
accurate distance and reddening estimates and to provide
a photometric abundance indicator for open clusters.

Motivated by the continuing discussion of open cluster
distances, we have undertaken a careful reexamination of
all the ingredients of the MS-fitting technique, revising
and extending our previous formulation \citep{pin98} and
taking advantage of new calibrations from Hipparcos and
from improvements to our stellar evolution models.  Our
first paper \citep[][hereafter Paper I]{paper1} discussed
the construction of the isochrones and demonstrated that
they provided a good match to the mass/luminosity relation
for the components of the eclipsing binary vB~22 in the
Hyades.  In this paper, we perform additional external
checks on the metallicity and temperature scales adopted
for the Hyades, then use photometry to generate empirical
adjustments to the isochrone colors and bolometric
corrections.  We conclude by discussing the sensitivity
of the MS luminosity to metallicity for various available
color calibrations.  In subsequent papers of this series,
we will extend the calibration to fainter magnitudes,
provide a grid of calibrated isochrones appropriate for
open clusters, and determine the effects of random and
systematic errors in the main-sequence fitting method.

\section{The empirical calibration}

\subsection{Data}

We adopted the Hyades as the zero point for the
MS-fitting method for a number of reasons.  The cluster
has a negligible reddening $E(B - V) = 0.003 \pm 0.002$
\citep{cra75,tay80} and excellent membership information
from proper motions and radial velocities.  Although the
cluster's size is a significant fraction of its distance,
the Hipparcos parallaxes and proper motions allow the
determination of the distances to individual Hyades
members to a precision of $\sim 2$\% \citep{bhz01},
with reasonable assumptions about the cluster's internal
motions.  The average distance, taken as that of the
cluster's center of mass ($m - M = 3.33 \pm 0.01$,
or $d = 46.34 \pm 0.27$ pc) from \citet{per98}, is not
altered significantly by correlations of the Hipparcos
parallaxes on small angular scales \citep{ng99b,bhz01}.
The most significant problem is that the Hipparcos data
do not extend below about $M_V = 8$, insufficient for
calibrating isochrones for the lower MS.  This will be
corrected by making use of data in other clusters such as
the Pleiades or Praesepe, as discussed in a future paper.

We chose as our Hyades sample a set of stars identified
as members of the cluster by \citet{per98}, excluding any
star that is a known or suspected binary.  Most of these
are included in the high-resolution spectroscopic sample
of \citet[][hereafter PSC]{paulson03}.

To construct a calibrating set of photometry for the
Hyades, we inspected the numerous $UBV$ photometric
studies and selected those which were in mutual agreement.
For $B$ and $V$, we selected stars from \citet{jk55},
\citet{men67}, and the Hipparcos mission (ESA 1997).
In $V$, we also had photometry from Weis and collaborators
\citep{uw77,weis79} that was also used for (Kron) $V
- I_{\rm K}$ colors (these will be discussed below).
The Hipparcos catalog provides individual errors for stars,
and the quoted errors in the Johnson and Mendoza papers
respectively are 0.004 and 0.01 magnitudes.  The dispersion
in $B - V$ between these three samples is consistent with
the quoted errors, and the zero-point difference for stars
in common is negligible (of order 0.001 mag or less).

The Hipparcos catalog lists $V$ magnitudes for all
of the stars in our sample; all but 32 of these are
derived from ground-based photometry.  There is a small
(0.012 mag) zero-point offset between the space-based
Hipparcos $V$ magnitudes and the Johnson/Mendoza data,
with dispersions of 0.013 and 0.009 mag respectively;
the difference is in the sense that the ground-based data
are fainter.  There are only six stars in common between
Weis and Johnson, with a formal dispersion of 0.015
mag and a zero-point offset of 0.005 (Johnson fainter),
which is not statistically significant.  We choose the
ground-based data for our zero-point, adding 0.012 mag to
the Hipparcos space-based $V$ magnitudes, and averaged the
results assuming an error of 0.01 for Hipparcos, Mendoza,
and Johnson and 0.015 for Weis.

The apparent $V$ magnitudes were converted into absolute
$M_V$ magnitudes using the individual distance moduli from
\citet{bhz01}.  The errors for $M_V$ are larger than for
$V$ (or $V$ used in the construction of $V - K_{\rm s}$
colors), since the error in the distances are included and
these errors are usually larger than those in $V$ alone.

We used two sources for the K band data: \citet{car82},
which is on the CTIO-CIT system \citep{fpam78,elias82},
and data from the 2MASS All Sky release.  The latter is on
the ``short-$K$'' ($K_s$) system \citep{persson98,car01}.
We converted the \citet{car82} $K$ magnitudes to $K_s$
using the transformation in \citet{car01} and averaged
these with the 2MASS values, using Carney's error estimate
of 0.011 mag for each star.  The error in the $V-K$
color was obtained by adding the errors in $V$ and $K_s$
in quadrature.

The merger of the $I$-band data for the clusters was
somewhat more involved than for $BVK$, largely because of
the need to transform different $I$-band systems (Johnson,
Kron/Eggen, and Cousins) onto the Cousins system.  There
were three main sources of Johnson data with a total of 60
stars: \citet{men67}, \citet{jmi66}, and \citet{jmm68}. In
addition there were six stars from \citet{ca79}.

Weis and collaborators \citep{uw77,weis79,wu82,uwh85,wh88}
have $V-I$ data in the Kron system for 23 stars.
There were also 11 stars with Johnson $V-I$ and accurate
$R-I$ colors from \citet{eggen82}, which are on the Kron
system \citep{eggen75}; we found that although the Eggen
and Johnson $R-I$ colors could be accurately transformed
onto the same system, the $V-I$ colors from Eggen had
larger errors.  This is probably because Eggen's $V$
magnitudes were not obtained at the same time as $R - I$,
so the errors in $V - I$ are much greater than in $R - I$.
Only five stars in the sample have $V - I$ colors in both
the Kron and Johnson systems.  We were unable to achieve
a good transformation between the merged colors and those
in \citet{tj85}, so we ignored their measurements.

The Johnson data were transformed onto the Cousins
system using the prescription in \citet{bessel79},
while the Weis and Eggen data were transformed onto the
Cousins system using the cubic formula in \citet{bw87}.
The average difference between the transformed Johnson and
Weis/Eggen colors are 0.009 and 0.004 mag respectively,
with a dispersion of 0.012 and 0.013 mag.  These error
estimates are only approximate because the number
of stars with multiple $V - I$ errors was small:
the bright ($V \lesssim 8$) stars in the Hyades 
mostly have Johnson photometry while the faint stars 
have Kron photometry and the magnitude range of
overlap is limited.  We therefore assumed that the
colors on both systems were correctly transformed onto
the Cousnins system, and did
did not apply any further zero-point or slope offsets to the
transformed $V-I$ data. 

The list of merged photometry is presented as Table 1.
The Hipparcos designation for each star is in the first
column, while the second lists other names.  The remaining
columns show the visual magnitude and colors, where the
errors represent either quoted errors in the various
datasets if only one measurement is available, or the
standard deviation if more than one datum was averaged.

For completeness, we have also included the $J - K$ and
$H - K$ colors for each star, even though they are not
otherwise discussed in this paper (in particular, the
metallicity sensitivity of $J - K$ is smaller than for
$V - K$, and there is very little variation in the values
of $H - K$ for main-sequence stars).  Note that several of
the brightest stars have large errors in the infrared colors
because they are saturated in the 2MASS survey.  We plan
to discuss the full set of infrared colors and their usefulness
for reddening and metallicity determinations in a subsequent
paper.

\subsection{Comparing different color calibrations}

In Paper I, we described the construction of an isochrone
for the Hyades and demonstrated that it compared favorably
to the masses and luminosities of the components of the
Hyades eclipsing binary vB~22.  The parameters for the
Hyades depend somewhat on the helium abundance assumed
for the Hyades and on the adopted solar model used to
calibrate the mixing length parameter for convection.
The isochrone used here is the same as in Paper I, namely
constructed from models ignoring microscopic diffusion
with $Y = 0.273$ and $\alpha = 1.72$.  The age of the
Hyades was taken as 550 Myr, appropriate for models lacking
convective overshoot \citep[e.g.,][]{per98}.  The models
used the solar abundance mix of \citet{gn93}, and scaled
to a Hyades abundance of [Fe/H] $= +0.13 \pm 0.01$ (PSC),
using the Sun's relative abundances.  The PSC average
abundance is quite close to that found by \cite{bf90};
see \citet{per98} for a summary of previous abundance
estimates in the Hyades.  The input physics is similar to 
that used in the recent \citet{ykd03}
$Y^2$ isochrones for stars of solar mass and above.  
Our usage of the \citet{scv95} equation of state for lower mass 
main sequence stars does cause intrinsic differences
between our isochrones and theirs for stars below 0.8 
solar masses.

In Figures 1--3 we compare the theoretical isochrone to
the Hyades photometry in Table 1, showing the effect of
using several available color-temperature calibrations.
Our base case, displayed as the solid line in these
figures, uses the corrected color-temperature relation
in \citet{lejeune98}, which employed multicolor data to
adjust theoretical flux distributions based on Kurucz
and \citet{ah95} atmospheres.  This isochrone matches
the photometry reasonably well, although systematic
departures from the data are readily apparent especially
for the reddest stars.\footnote{The various available
color-temperature relations differ the most for stars
cooler than about 4500 K.  The empirically corrected
Lejeune et al.\ (1998) relation, for example, is nearly
identical to that presented by \citet{flower96}, but is bluer
by nearly 0.1 mag at 4000 K.} The three color-magnitude
diagrams (CMDs) are in $M_V$ and $B - V$, $V - I_C$, and
$V - K_s$, respectively.  The absolute magnitudes were
generated from the $V$ magnitudes and individual kinematic
parallaxes from \citet{bhz01} adjusted to a common distance
modulus of $m - M = 3.33 \pm 0.01$ \citep{per98}.

Also shown on Figures 1--3 as a short-dashed line is
the same isochrone generated with the color-temperature
relation in \citet{alonso95,alonso96};  equivalently,
the long-dashed line displays the isochrone made
with the \citet{lejeune98} calibration {\it before}
their application of empirical corrections to the
theoretical colors.  Here, we will refer to the latter as
the uncorrected \cite{lejeune98} calibration, though note
that their terminology is different.

\subsection{Isochrone effective temperatures}

We have just shown that a single theoretical isochrone
will produce different loci in the color-magnitude diagram
when different color-temperature relations are applied.
Before correcting the colors to match the Hyades photometry,
it is necessary to do one additional test: namely to verify
that the theoretical quantities generated by the 
stellar evolution models ($L$, $T_{\rm eff}$, $\log g$) are
reasonably close to that of the actual cluster,
since any errors will be washed away by forcing the
isochrone colors to match the photometry.  The recent
spectroscopic temperature determinations by PSC allow
this test to be done with precision.

In both
theoretical models and direct parallax measurements,
high metallicity stars appear fainter at a fixed color
than low metallicity stars.  This is partially a stellar
interiors effect;  increased metal abundance makes stars
of fixed mass slightly fainter and significantly cooler,
moving them above the MS locus of lower-metallicity stars.
Increased line blanketing in more metal-rich stars also
makes them appear redder at fixed effective temperature.
Increasing the helium abundance will make stars appear
systematically fainter at fixed effective temperature
or color;  however, the spectral energy distribution and
color-temperature relationship is largely insensitive to
the helium abundance.

We now proceed to examine the match between the temperature
and gravities in the isochrone to parameters derived by
PSC.  They determined effective temperatures for each star
by the requirement that the \ion{Fe}{1} abundances from
several lines were independent of excitation potential.
Gravity was found from requiring ionization equilibrium
that yields the same abundance for \ion{Fe}{1} and
\ion{Fe}{2} lines.

In Figure 4, we display (top panel) the difference
between the PSC temperatures and those from the isochrone
with the \citet{lejeune98} color calibration, namely
the isochrone plotted as a solid line in Fig.\ 3.
The difference is in the sense of isochrone
{\it minus} spectra, and the isochrone values for each
star were found by looking up the effective temperature at
each $M_V$. These temperatures are therefore independent of
systematic errors in the color calibration, discussed below
in $\S$ 3.  The lower panel shows the difference in $\log
g$ constructed in a similar manner.  The circular points
show stars that are in our Table 1, while the triangles
show other stars in PSC;  these were not included in our
sample either because they may be binaries or because there
was accurate photometry but not accurate luminosities.

Overall, the agreement between the isochrone parameters and
the spectroscopically derived values is remarkably good,
with average differences of only $-24$ K in $T_{\rm eff}$
and $+0.02$ in $\log g$.  There are, however, noticeable
trends with temperature, in that the PSC temperatures are
higher than the isochrone values for the hottest stars
in their sample; for these stars, their adopted gravities
are somewhat too low.  In combination, their temperatures
and gravities imply masses about $0.2 M_\odot$ above the
isochrone values, or radii that are about 5\% too large
at fixed mass compared to the isochrones.

In their Table 7, PSC provide coefficients showing the
sensitivity of the derived abundances to changes in
temperature and gravity, from which we can rescale their
[Fe/H] to the temperatures and gravities of the isochrones
(we adopt their microturbulence values without change).
The coefficients were tabulated for two stars at the
extremes of the temperature distribution, from which
we linearly interpolated to determine the sensitivity
of the abundances at any intermediate temperature.
Figure 5 displays (top panel) the PSC values of [Fe/H]
as a function of isochrone effective temperature, where
the symbols are the same as in the previous figure.
The rescaled abundances are displayed in the lower panel.

The mean derived spectroscopic abundance using the
rescaled abundances is $\langle{\rm [Fe/H]}\rangle =
+0.12 \pm 0.01$.  This is in excellent agreement with the
original assumed abundance of +0.13, showing the locus of
the isochrone in the theoretical plane is fully consistent
with spectroscopic abundance data to a high degree of
accuracy.\footnote{Even though we started with the PSC
value in constructing the isochrone, the agreement we find
does not amount to a circular argument.  If we had adopted
(say) [Fe/H] $= 0.0$ for the isochrone, that isochrone
would have been too faint at the Hipparcos distance to
the cluster.  The temperatures derived from $M_V$ would
be therefore be too high, which would have required
the PSC abundances to be increased significantly.} The
dispersion in the rescaled abundances is slightly smaller
than that obtained with the spectroscopic temperatures
and gravities, the number of outliers is reduced,
and any underlying trends in [Fe/H] with $T_{\rm eff}$
are small.  We therefore conclude that the models in the
theoretical plane are a reasonable representation of the
actual cluster.

\subsection{A recap}

To summarize, our argument so far has been thus: (1)
Solar models are in excellent agreement with the stringent
tests possible with helioseismic data \citep{bpb00}, so
models for Hyades stars of similar effective temperatures
should be similarly well constrained; (2) The theoretical
mass/luminosity relationship is in good agreement with
the values obtained from the eclipsing binary vB~22 for a
reasonable Hyades helium abundance (paper I); (3) Available
color-temperature relations applied to the same isochrone
do not match the photometry of the Hyades ($\S$ 2.2); (4)
The spectroscopic temperatures of Hyades are consistent
with the luminosity / effective temperature relationship
in the models ($\S$ 2.3).

We therefore believe that the isochrones in the theoretical
luminosity-effective temperature plane are well matched
to the Hyades.  In other words, the stellar interiors
models are not the cause of the mismatch in shape between
the isochrones and the Hyades photometry in Figs.\ 1--3.
The deviations for any one isochrone are usually about
the same size as the differences between published color
calibrations, which suggests that we would be justified
in applying modest empirical corrections to the isochrones.

We now proceed to define the empirical locus of the Hyades
in the $M_V$-color plane and obtain a set of corrections to
our base color-temperature relationship \citep{lejeune98}.
We begin by examining the internal consistency of effective
temperatures derived from published color-temperature
relationships for the same star but with multiple colors.
We then define empirical color-color relationships for
the Hyades, tying the other colors to $B-V$, and identifying
the locus of the cluster in the ($M_V$, $B-V$) plane.
When combined with the color-color relationships, this
defines the locus of the cluster in each of the color indices
that we have included in this paper (namely $B-V$, $V-I_C$,
and $V-K_s$;  other colors will be treated in subsequent
papers).  Finally, we use the isochrone $M_V$, $T_{\rm
eff}$ relationship to convert these empirical corrections
as a function of $M_V$ into empirical corrections to the
color-temperature relationship.

\subsection{The internal consistency of the
color-temperature relationships}

If the metallicity and reddening of a cluster are known,
a self-consistent set of color-temperature relationships
should yield the same temperatures for each star regardless
of which color is used to derive the temperature.
In Figure 6, we compare the $T_{\rm eff}$ estimates for
each star in Table 1 using the \citet{lejeune98} color
calibration.  The horizontal axis displays the isochrone
effective temperature derived from $M_V$; note that the
temperatures extend to higher values than in Figures 4
and 5 because the PSC survey only included stars cooler
than about $T_{\rm eff} = 6300$ K.  The top panel shows
the difference between the temperature estimated from $V -
I_C$ and that from $B - V$;  the middle and lower panels,
respectively, compare the temperature from $V - K_s$ to $B
- V$ and $V - I_C$, again as a function of the temperature
from $M_V$.  Errors in the temperatures are propagated
from errors in the photometry.

While the mean temperature differences are nearly zero
(the largest is $\langle\Delta T\rangle = -60$ K for
the data in the top panel), there are clear systematic
differences as a function of temperature which can be as
large as 300 K.  This indicates that there are temperature
ranges where the \citet{lejeune98} color calibration is not
internally consistent, at least with respect to the Hyades
multicolor photometry.  These small inconsistencies may
have arisen because the color transformation was derived
by comparison to an ensemble of field stars with various
metallicities;  the (generally) small differences noted
here would probably not have been apparent in the
Lejeune et al. analysis.  

We also explored internal inconsistencies using
other color-temperature relationships, and found
similar results.  We therefore
conclude that the color-temperature calibrations need
to be modified not only to remove systematic errors
in distances and photometric metallicity, but also for
internal consistency.  Recently, \cite{gs03} have compared
multicolor photometry in many clusters to a number of
color-temperature relations and have come to a similar
conclusion.

\subsection{Correcting the color-color relationship}

The next step is to correct the color-color relationship.
There were several reasons why we performed this correction
rather than adjusting the isochrones in the ($M_V$, color)
planes individually.  First, the errors in the colors are
typically smaller than the errors in $M_V$; the latter,
of course, arise from parallax errors and are much larger
than the error in the $V$ magnitudes.  Second, correcting
the color-color relation necessarily produces the desired
result that the isochrone color-temperature relation be
internally consistent (i.e., each color yields the same
temperature for a star, at least to the limits set by
photometric errors).

The smallest observational errors are in $B - V$, so
we chose to define corrections in the other colors as a
function of $B-V$.  The method is illustrated in Figure 7.
The top panel shows the color difference $\Delta(V - I_C)$
between the isochrone and the photometry for each star
in Table 1, plotted against the observed $B - V$ color.
The lower panel shows the same, but for $V - K_s$.
The isochrone employed was the same base case (above),
namely for [Fe/H] $= +0.13$ and the \citet{lejeune98}
color calibration;  we have neglected systematic errors
arising from the small uncertainty in the metallicity of
the Hyades.  The color differences are in the sense of
(photometry -- isochrone), so positive values indicate
that the star is redder than the isochrone.  The isochrone
colors $V - I_C$ and $V - K_s$ were computed by finding
the effective temperature at each $B - V$, then looking
up the other colors at that temperature.  The error bars
on each point include both the observational error in the
photometry and the error in the predicted color arising
from errors in $B - V$.

The smooth line in each panel of Figure 7 represents the
color correction to $V - I_C$ or $V - K_s$ as a function
of $B - V$ (specifically the $B - V$ color of the Hyades
stars).  The line was constructed by finding the average
value of $\Delta(V - I_C)$ or $\Delta(V - K_s)$ in a moving
box containing 3--5 points sorted by increasing $B - V$.
The line was then smoothed by averaging each point with 
the linear interpolation of adjacent points blueward and
redward.

These corrections would generate $V - I_C$ and $V - Ks$
colors that are consistent with the Hyades photometry
once the isochrone generates the observed $B - V$
colors at the adopted temperature scale.  By defining
the color corrections in this way, we assume that the
$V$-band bolometric corrections in the isochrones are
good;  $\Delta(V - I_C)$ is then taken as an adjustment
to $BC(I_C)$ and applied to the calculated $I$ magnitude
of the isochrone (and equivalently for $B$ and $K_s$).

The final step is to make the isochrone generate the
right $B - V$, as illustrated in Figure 8, which plots the
quantity $\Delta(B - V)$ against the isochrone effective
temperature as derived from $M_V$.  Here the smoothing was
done in a moving box of width 200 K centered at intervals
of 100 K, with adjacent pairs averaged.

Applying these corrections to the colors at each effective
temperature defines the empirically calibrated isochrone
for the Hyades, which is tabulated along with the color
corrections in Table 2.  The range of colors over which
the correction was performed was limited by the following
considerations.  The Hyades turnoff is around $B - V = 0.2$
and the magnitude-limited Hipparcos sample has few stars
with $B - V > 1.3$.  There are only a few stars with $B -
V \leq 0.4$ so the correction is not well defined for the
bluest colors.  Therefore the most reliable color range
for the corrections is $0.4 \leq B - V \leq 1.3$.

\section{The color calibrations and metallicity}

The process of empirically correcting the isochrones
to match the Hyades necessarily erases many potential
systematic errors in the input physics, metallicity
scale, or the computation of isochrone colors.  On the
positive side, the isochrone now matches the shape of the
Hyades main sequence to high precision and (on average)
will yield the same effective temperature for the all
the various colors.  We are thus prepared to derive
more accurate distances and metallicities for other open
clusters relative to the set of parameters assumed for the
Hyades: ${\rm [Fe/H]} = +0.13$, scaled solar abundances,
$Y = 0.273$, $\alpha = 1.72$, $\langle (m - M)_0\rangle =
3.33$, etc.

There is, however, one more step before we can examine
other clusters in detail, namely measuring the sensitivity
of the isochrone luminosity to metallicity in the several
colors.  For this, we return to the individual color
calibrations discussed above.  The isochrone absolute
magnitude $M_V$ depends both on the luminosity of the
isochrone in the theoretical plane and on the metallicity
sensitivity of the adopted color-temperature relations.
We have defined our empirical isochrone with respect
to the \citet{lejeune98} color-temperature relation, and
assume that the color corrections displayed in Table 2
are independent of metallicity.  As a result, we will
be measuring cluster distances on this system, and wish
to compare this to the available alternatives.

In Figure 9, we illustrate the sensitivity of the
main-sequence luminosity to metallicity for the three
color-temperature relations we explored in $\S$ 2.2.
The top panel plots the quantity \[ \Delta M_V(B - V) =
\left. \langle
      M_V({\rm [Fe/H]})
    - M_V({\rm [Fe/H]} = +0.13)
\rangle \right|_{B - V = 0.4 - 1.0}, \] namely the average
difference in absolute magnitude between the isochrone at
any particular metallicity and that for ${\rm [Fe/H]} =
+0.13$, computed over the range $0.4 \leq (B - V) \leq 1.0$
at constant $B - V$.   The points are at intervals of 0.1
in metallicity, from ${\rm [Fe/H]} = -0.3$ to $+0.2$, and
were computed by finding the average difference in $M_V$
between the stellar values in Table 1 and the isochrone
for that metallicity.  The errors on the individual
points are on the order of 0.01 mag.  The values for
the Hyades isochrone employing the \citet{lejeune98}
color calibration are shown as filled points;  the
adjacent straight line is a linear least-squares fit
to those points.  The uncorrected \citet{lejeune98} and
\citet{alonso95,alonso96} color calibrations are displayed
as open circles and triangles, respectively.  The cross at
${\rm [Fe/H]} = +0.13$ and $\Delta M_V = 0.0$ shows where
the isochrones should go if they did not require empirical
corrections to the color-temperature relation (i.e.,
if they matched the Hyades photometry {\it a priori}.)

The center and lower panels, respectively, plot the mean
difference at fixed $V - I_C$ and $V - K_s$, computed over
the same color interval in $B - V$.

The value of $\Delta M_V$ enters into the computation of
cluster distances in several different ways.  The slope
of $\Delta M_V$ as a function of [Fe/H] is a measure of
distance errors that arise from errors in the assumed
or measured metallicity of a cluster.  The average slope
over the range $-0.3 \leq {\rm [Fe/H]} \leq +0.2$ is shown
for the three representative color calibrations in Table 3
(the relationship is probably not precisely linear, having
a somewhat steeper dependence on metallicity above (${\rm
[Fe/H]} = 0.0$).  All the color calibrations have nearly
the same dependence on $M_V$ at fixed $B - V$, with
an average slope of 1.42 mag per dex in metallicity.
The metallicity dependence is less steep at fixed
$V - I_C$ or fixed $V - K_s$, and we note that the
\citet{alonso95,alonso96} calibration produces a markedly
shallower metallicity dependence of the main-sequence
luminosity in at constant $V - I_C$ 
than the other two calibrations.

While the various calibrations generally agree on the
slope $d(\Delta M_V) / d{\rm [Fe/H]}$, they have very
different zero points.  Since these calculations are
all done for the same input physics and metallicity
($\rm [Fe/H] = +0.13$), one would necessarily derive
different distances to the Hyades using the various color
calibrations.  Forcing the distances to be the same for
the various colors yields a photometric metallicity index
for a cluster \citep[e.g.,][]{pin98}.

The last two columns of Table 3 display the value of [Fe/H]
that would be found for the various color calibrations,
and the relative distance with respect to the assumed
value of 3.33 for the Hyades.

\section{Summary and discussion}

In this paper, we have continued our effort to improve
the calibration of main-sequence isochrones for use
in determining the distances and metallicities of open
clusters.  We first identified a set of single Hyades
stars with quality multicolor photometry, then verified
that the isochrone luminosity-temperature relationship is
in agreement with recent spectroscopic determinations.
We then described a method for computing empirical
corrections to the color-temperature relationship in
the isochrone, and explored the sensitivity of the
main-sequence luminosity to metallicity using several
available color-temperature calibrations (see also
VandenBerg and Clem 2003 for a parallel discussion of
these issues).  In subsequent papers, we will extend
the empirical calibration to the lower main sequence via
photometry in other clusters, determine the accuracy of
photometric metallicity indicators, explore the effects
of reddening, and derive ages for young systems via an
analysis of the pre-main sequence / main sequence boundary.

The principal result of this paper was that the theoretical
and spectroscopic (PSC) luminosity-temperature and
luminosity-radius (i.e., $L$, $\log g$) relations are
in fairly good agreement (Fig.\ 4) except for stars with
$T_{\rm eff} \ga 5800$.  The spectroscopic $\log g$ values,
if real, would imply large errors in the model radii that
would cause much larger differences in $T_{\rm eff}$ than
seen (assuming that the mass/luminosity relationship
is correct).  They are also very unlikely given the
similarity of the stars to the Sun, which should imply
that the solar calibration produces reasonable radii
for models of Hyades stars.  We have verified that the
dispersion in all of the abundances measured by PSC
is significantly reduced when the theoretical $\log g$
values are used in place of the spectroscopic ones (a
more detailed analysis will appear in a future paper).
Paulson et al.\ note similar trends with respect to
temperature scales employed by \citet{apl99}, which
was based on a different set of theoretical isochrones.
We suggest that there may be potential degeneracies between
the spectroscopic temperature and gravity measurements.
The spectroscopic $T_{\rm eff}$ and $\log g$ values are
derived from Saha/Boltzmann considerations, and the fitting
procedure used by PSC may compensate for errors in the
ionization balance ($\log g$) by changing the excitation
balance ($T_{\rm eff}$).  This idea could be tested by
seeing what temperature scale resulted if the gravities
were taken from the theoretical isochrones rather than
treating them as a free parameter.  In any case, the
overall agreement in between a high-quality spectroscopic
data set and the isochrones in the theoretical plane
provides strong support for the hypothesis that the model
temperature scale is reasonable.

The main goal of this effort is to reduce the theoretical
and calibration uncertainties in the theoretical models
using the constraints provided by the detailed morphology
of the main sequence in various clusters.  As outlined in
the Introduction, however, we have long been interested in
the discrepancy between the Hipparcos distances to clusters
\citep{esa97} as subsequently analyzed by \citet{rob99},
\cite{vl99}, and others \citep[see][]{pin98,sod98}.
Various attempts to explain away the discrepancies as
arising from, say, strange helium abundances \citep{bel98}
or stellar activity \citep{per03}, whether true or not
(and worth exploring further), do raise an important
issue: accurate calibration of isochrone physics and
colors is an instrinsically difficult exercise, whereas
parallaxes should be easier in principle (though perhaps
not in practice).

In our view, the issue of the distance discrepancies
arising from the 1997--1999 Hipparcos parallaxes is
largely settled.  Several papers have pointed out that
small-scale errors in the parallaxes are effectively random
zero-point errors over the whole sky, with an r.m.s. error
of about 0.5 mas.  The Hyades are less affected than other
Hipparcos clusters because the Hyades subtend far larger
than 1 degree, allowing the small-scale errors to average
out \cite{ng99b}.  The Pleiades are strongly affected
because the bright stars are quite centrally concentrated
(17 Hipparcos stars per square degree at the cluster
center) and so they share the same zero-point error, which
amounts to $+1$ mas at that point in the sky.  Recently,
\citet{makarov02} has devised a method to recalculate
parallaxes from the Hipparcos Intermediate Astrometric
Data, free of the small angular scale correlated errors
which are an artifact of the ``great circle abscissa
method" data reductions used in the \citet{esa97}
distances.  His new Pleiades distance is in agreement
with the distance from main-sequence fitting, and also
with Pleiades distances derived from kinematic parallaxes
\citep{ng99b} and new ground-based results \citep{gate00}.

\acknowledgements

The work reported here was supported in part by the
National Science Foundation, under grants AST-9731621
and AST-0206008 to the Ohio State University Research
Foundation.

\clearpage \begin{deluxetable}{llrccccccccccccccc}
\tablewidth{0pt}
\tabletypesize{\scriptsize}
\rotate
\setlength{\tabcolsep}{0.02in}
\tablecaption{Merged Photometry in the Hyades}
\tablehead{
  \colhead{Hipp} &
  \colhead{Other ID\tablenotemark{a}} &
  \colhead{$V$} &
  \colhead{$\sigma(V)$} &
  \colhead{$M_V$} &
  \colhead{$\sigma(M_V)$} &
  \colhead{$B - V$} &
  \colhead{$\sigma(B-V)$} &
  \colhead{$V-I_C$} &
  \colhead{$\sigma(V-I_C)$} &
  \colhead{$K_s$} &
  \colhead{$\sigma(K_s)$} &
  \colhead{$V-K_s$} &
  \colhead{$\sigma(V-K_s)$} &
  \colhead{$J - K_s$} &
  \colhead{$\sigma(J-K_s)$} &
  \colhead{$H-K_s$} &
  \colhead{$\sigma(H-K_s)$}
}
\startdata
13806 & vB~153    &  8.915 & 0.007 & 5.856 & 0.026 & 0.856 & 0.003 &   0.871 &   0.016 & 6.905 & 0.022 & 2.010 & 0.023 & 0.478 & 0.025 & 0.092 & 0.023 \\
13834 & vB~154    &  5.800 & 0.010 & 3.222 & 0.019 & 0.410 & 0.003 &   0.468 &   0.011 & 4.778 & 0.020 & 1.022 & 0.022 & 0.530 & 0.198 & 0.037 & 0.035 \\
13976 & G76-49    &  7.970 & 0.010 & 6.131 & 0.019 & 0.926 & 0.015 & \nodata & \nodata & 5.841 & 0.024 & 2.129 & 0.026 & 0.482 & 0.022 & 0.110 & 0.043 \\
14976 & LTT~11045 &  8.162 & 0.010 & 5.156 & 0.025 & 0.732 & 0.017 & \nodata & \nodata & 6.557 & 0.027 & 1.605 & 0.029 & 0.368 & 0.026 & 0.098 & 0.031 \\
15563 & G79-28    &  9.650 & 0.010 & 7.187 & 0.028 & 1.130 & 0.015 & \nodata & \nodata & 6.881 & 0.022 & 2.769 & 0.024 & 0.664 & 0.024 & 0.074 & 0.052 \\
15720 & LP~355-64 & 11.030 & 0.010 & 8.488 & 0.044 & 1.431 & 0.004 & \nodata & \nodata & 7.415 & 0.022 & 3.615 & 0.024 & 0.834 & 0.023 & 0.175 & 0.039 \\
16529 & vB~4      &  8.890 & 0.010 & 5.765 & 0.032 & 0.845 & 0.002 &   0.856 &   0.016 & 6.907 & 0.016 & 1.983 & 0.019 & 0.467 & 0.014 & 0.111 & 0.010 \\
17766 & G7-15     & 10.850 & 0.010 & 8.033 & 0.048 & 1.340 & 0.006 &   1.558 &   0.016 & 7.509 & 0.016 & 3.341 & 0.019 & 0.764 & 0.020 & 0.115 & 0.030 \\
18170 & vB~6      &  5.970 & 0.010 & 2.843 & 0.030 & 0.341 & 0.004 &   0.389 &   0.016 & 5.091 & 0.018 & 0.879 & 0.021 & 0.163 & 0.022 & 0.070 & 0.041 \\
18322 & L8, G7-34 & 10.120 & 0.010 & 6.779 & 0.053 & 1.070 & 0.008 &   1.124 &   0.020 & 7.574 & 0.021 & 2.546 & 0.023 & 0.607 & 0.020 & 0.118 & 0.024 \\
18327 & vB~7      &  8.987 & 0.010 & 5.927 & 0.037 & 0.895 & 0.002 &   0.899 &   0.011 & 6.910 & 0.021 & 2.077 & 0.023 & 0.503 & 0.019 & 0.128 & 0.027 \\
19082 & L12       & 11.415 & 0.007 & 8.007 & 0.071 & 1.348 & 0.004 &   1.577 &   0.020 & 8.107 & 0.026 & 3.308 & 0.027 & 0.782 & 0.034 & 0.152 & 0.031 \\
19098 & L10       &  9.310 & 0.015 & 6.032 & 0.044 & 0.893 & 0.005 &   0.896 &   0.016 & 7.254 & 0.022 & 2.056 & 0.027 & 0.494 & 0.030 & 0.111 & 0.039 \\
19148 & vB~10     &  7.850 & 0.006 & 4.449 & 0.036 & 0.591 & 0.003 &   0.607 &   0.016 & 6.446 & 0.024 & 1.404 & 0.025 & 0.307 & 0.025 & 0.086 & 0.038 \\
19207 & L15       & 10.485 & 0.007 & 7.163 & 0.051 & 1.180 & 0.008 &   1.268 &   0.020 & 7.665 & 0.021 & 2.820 & 0.022 & 0.700 & 0.028 & 0.146 & 0.051 \\
19263 & BD+16:558 &  9.940 & 0.010 & 6.628 & 0.044 & 1.005 & 0.012 & \nodata & \nodata & 7.511 & 0.026 & 2.429 & 0.028 & 0.594 & 0.027 & 0.088 & 0.044 \\
19316 & L14       & 11.277 & 0.006 & 7.871 & 0.074 & 1.328 & 0.003 &   1.558 &   0.016 & 8.046 & 0.027 & 3.231 & 0.028 & 0.772 & 0.033 & 0.125 & 0.045 \\
19441 & BD+8:642  & 10.100 & 0.010 & 7.361 & 0.045 & 1.198 & 0.006 &   1.278 &   0.016 & 7.264 & 0.020 & 2.836 & 0.022 & 0.648 & 0.016 & 0.076 & 0.021 \\
19781 & vB~17     &  8.460 & 0.007 & 4.947 & 0.046 & 0.696 & 0.004 &   0.720 &   0.011 & 6.805 & 0.009 & 1.655 & 0.012 & 0.391 & 0.025 & 0.091 & 0.021 \\
19786 & vB~18     &  8.060 & 0.007 & 4.735 & 0.046 & 0.640 & 0.002 &   0.677 &   0.011 & 6.535 & 0.009 & 1.525 & 0.012 & 0.322 & 0.013 & 0.073 & 0.017 \\
19789 & vB~16     &  7.057 & 0.006 & 3.283 & 0.037 & 0.420 & 0.003 &   0.451 &   0.016 & 6.005 & 0.021 & 1.052 & 0.022 & 0.202 & 0.017 & 0.077 & 0.023 \\
19793 & vB~15     &  8.067 & 0.020 & 4.820 & 0.039 & 0.658 & 0.003 &   0.692 &   0.011 & 6.516 & 0.020 & 1.551 & 0.028 & 0.361 & 0.023 & 0.083 & 0.018 \\
19796 & vB~19     &  7.130 & 0.010 & 3.846 & 0.038 & 0.512 & 0.003 &   0.560 &   0.016 & 5.862 & 0.017 & 1.268 & 0.020 & 0.266 & 0.013 & 0.053 & 0.017 \\
19808 & vA~68     & 10.725 & 0.050 & 7.424 & 0.085 & 1.205 & 0.005 &   1.288 &   0.011 & 7.680 & 0.009 & 3.045 & 0.050 & 0.757 & 0.025 & 0.140 & 0.013 \\
19834 & vA~72     & 11.557 & 0.015 & 8.197 & 0.115 & 1.373 & 0.006 &   1.630 &   0.011 & 8.141 & 0.009 & 3.416 & 0.018 & 0.819 & 0.024 & 0.166 & 0.021 \\
19877 & vB~20     &  6.321 & 0.007 & 2.987 & 0.033 & 0.399 & 0.004 &   0.459 &   0.016 & 5.340 & 0.009 & 0.981 & 0.011 & 0.188 & 0.030 & 0.075 & 0.020 \\
19934 & vB~21     &  9.147 & 0.006 & 5.625 & 0.043 & 0.814 & 0.002 &   0.825 &   0.011 & 7.275 & 0.022 & 1.872 & 0.023 & 0.426 & 0.020 & 0.101 & 0.020 \\
20130 & vB~26     &  8.625 & 0.007 & 5.323 & 0.045 & 0.743 & 0.003 &   0.762 &   0.016 & 6.921 & 0.016 & 1.704 & 0.017 & 0.365 & 0.012 & 0.079 & 0.012 \\
20146 & vB~27     &  8.464 & 0.015 & 5.138 & 0.045 & 0.717 & 0.004 &   0.749 &   0.011 & 6.793 & 0.009 & 1.671 & 0.018 & 0.347 & 0.017 & 0.085 & 0.015 \\
20219 & vB~30     &  5.591 & 0.007 & 2.331 & 0.035 & 0.279 & 0.003 &   0.338 &   0.011 & 4.861 & 0.009 & 0.730 & 0.011 & 0.149 & 0.033 & 0.074 & 0.045 \\
20237 & vB~31     &  7.467 & 0.006 & 4.200 & 0.034 & 0.566 & 0.004 &   0.622 &   0.016 & 6.124 & 0.018 & 1.343 & 0.019 & 0.271 & 0.013 & 0.060 & 0.024 \\
20261 & vB~33     &  5.266 & 0.007 & 1.879 & 0.037 & 0.223 & 0.003 &   0.272 &   0.011 & 4.689 & 0.016 & 0.577 & 0.017 & 0.061 & 0.046 & 0.284 & 0.282 \\
20349 & vB~35     &  6.801 & 0.006 & 3.329 & 0.039 & 0.436 & 0.004 &   0.513 &   0.016 & 5.762 & 0.020 & 1.039 & 0.021 & 0.208 & 0.017 & 0.024 & 0.015 \\
20350 & vB~36     &  6.807 & 0.006 & 3.457 & 0.035 & 0.441 & 0.004 &   0.498 &   0.016 & 5.746 & 0.016 & 1.061 & 0.017 & 0.198 & 0.015 & 0.048 & 0.010 \\
20357 & vB~37     &  6.611 & 0.006 & 3.160 & 0.038 & 0.406 & 0.004 &   0.475 &   0.016 & 5.610 & 0.009 & 1.000 & 0.011 & 0.189 & 0.019 & 0.071 & 0.031 \\
20480 & vB~42     &  8.854 & 0.010 & 5.338 & 0.048 & 0.758 & 0.002 &   0.786 &   0.016 & 7.121 & 0.016 & 1.733 & 0.019 & 0.410 & 0.011 & 0.076 & 0.011 \\
20485 & vB~173    & 10.499 & 0.020 & 7.485 & 0.072 & 1.237 & 0.003 &   1.368 &   0.011 & 7.481 & 0.009 & 3.018 & 0.022 & 0.669 & 0.019 & 0.153 & 0.039 \\
20491 & vB~44     &  7.185 & 0.007 & 3.541 & 0.042 & 0.452 & 0.003 &   0.521 &   0.016 & 6.092 & 0.016 & 1.093 & 0.017 & 0.193 & 0.015 & 0.049 & 0.010 \\
20492 & vB~46     &  9.117 & 0.006 & 5.731 & 0.054 & 0.865 & 0.003 &   0.871 &   0.011 & 7.153 & 0.009 & 1.964 & 0.011 & 0.466 & 0.011 & 0.110 & 0.016 \\
20527 & vA~294    & 10.901 & 0.020 & 7.673 & 0.075 & 1.288 & 0.002 &   1.465 &   0.009 & 7.755 & 0.009 & 3.146 & 0.022 & 0.720 & 0.021 & 0.139 & 0.021 \\
20542 & vB~47     &  4.800 & 0.010 & 1.526 & 0.038 & 0.156 & 0.003 &   0.179 &   0.016 & 4.400 & 0.010 & 0.400 & 0.014 & 0.147 & 0.268 & 0.099 & 0.210 \\
20557 & vB~48     &  7.141 & 0.006 & 4.010 & 0.037 & 0.521 & 0.004 &   0.576 &   0.016 & 5.871 & 0.020 & 1.270 & 0.021 & 0.277 & 0.020 & 0.075 & 0.025 \\
20563 & vB~174    &  9.995 & 0.005 & 6.740 & 0.064 & 1.058 & 0.003 &   1.101 &   0.011 & 7.525 & 0.009 & 2.470 & 0.011 & 0.630 & 0.013 & 0.112 & 0.020 \\
20567 & vB~51     &  6.971 & 0.006 & 3.461 & 0.047 & 0.443 & 0.004 &   0.517 &   0.016 & 5.880 & 0.009 & 1.090 & 0.011 & 0.167 & 0.014 & 0.043 & 0.016 \\
20635 & vB~54     &  4.220 & 0.010 & 0.846 & 0.034 & 0.137 & 0.003 &   0.163 &   0.011 & 4.077 & 0.470 & 0.404 & 0.470 & 0.016 & 0.549 & --0.013& 0.522 \\
20641 & vB~55     &  5.281 & 0.007 & 2.048 & 0.033 & 0.248 & 0.003 &   0.311 &   0.016 & 4.607 & 0.016 & 0.674 & 0.017 & 0.479 & 0.250 & 0.314 & 0.230 \\
20741 & vB~64     &  8.120 & 0.007 & 4.850 & 0.046 & 0.659 & 0.003 &   0.694 &   0.011 & 6.554 & 0.010 & 1.566 & 0.012 & 0.346 & 0.021 & 0.096 & 0.036 \\
20762 & vA 407    & 10.475 & 0.007 & 7.118 & 0.062 & 1.146 & 0.002 &   1.230 &   0.016 & 7.743 & 0.010 & 2.732 & 0.013 & 0.637 & 0.046 & 0.122 & 0.037 \\
20815 & vB~65     &  7.421 & 0.006 & 4.054 & 0.039 & 0.536 & 0.004 &   0.584 &   0.011 & 6.129 & 0.010 & 1.292 & 0.012 & 0.234 & 0.020 & 0.033 & 0.028 \\
20826 & vB~66     &  7.507 & 0.010 & 4.252 & 0.042 & 0.556 & 0.003 &   0.607 &   0.016 & 6.158 & 0.017 & 1.349 & 0.020 & 0.266 & 0.011 & 0.064 & 0.027 \\
20827 & vB~179    &  9.502 & 0.015 & 6.063 & 0.060 & 0.931 & 0.003 &   0.959 &   0.016 & 7.346 & 0.010 & 2.156 & 0.018 & 0.546 & 0.028 & 0.108 & 0.021 \\
20842 & vB~67     &  5.720 & 0.010 & 2.251 & 0.039 & 0.271 & 0.004 &   0.303 &   0.016 & 5.055 & 0.018 & 0.665 & 0.021 & 0.108 & 0.011 & 0.048 & 0.014 \\
20894 & vB~72     &  3.410 & 0.010 & 0.146 & 0.036 & 0.179 & 0.003 &   0.210 &   0.011 & 2.896 & 0.011 & 0.514 & 0.015 & 0.114 & 0.335 & 0.117 & 0.339 \\
20899 & vB~73     &  7.847 & 0.010 & 4.521 & 0.041 & 0.609 & 0.003 &   0.626 &   0.011 & 6.398 & 0.010 & 1.449 & 0.014 & 0.292 & 0.021 & 0.058 & 0.025 \\
20948 & vB~78     &  6.914 & 0.010 & 3.609 & 0.042 & 0.451 & 0.002 &   0.516 &   0.009 & 5.791 & 0.010 & 1.123 & 0.014 & 0.236 & 0.029 & 0.078 & 0.034 \\
20949 & vB~76     &  9.206 & 0.030 & 5.394 & 0.057 & 0.764 & 0.002 &   0.768 &   0.011 & 7.462 & 0.029 & 1.744 & 0.042 & 0.382 & 0.028 & 0.072 & 0.034 \\
20951 & vB~79     &  8.955 & 0.007 & 5.697 & 0.055 & 0.831 & 0.002 &   0.827 &   0.009 & 7.056 & 0.010 & 1.899 & 0.012 & 0.469 & 0.029 & 0.094 & 0.029 \\
20978 & vB~180    &  9.090 & 0.030 & 5.807 & 0.051 & 0.854 & 0.003 &   0.879 &   0.016 & 7.120 & 0.010 & 1.970 & 0.032 & 0.480 & 0.025 & 0.072 & 0.030 \\
21029 & vB~82     &  4.786 & 0.007 & 1.484 & 0.034 & 0.171 & 0.002 &   0.187 &   0.016 & 4.366 & 0.011 & 0.420 & 0.013 & 0.409 & 0.226 & 0.313 & 0.203 \\
21036 & vB~84     &  5.406 & 0.007 & 2.159 & 0.036 & 0.262 & 0.003 &   0.311 &   0.016 & 4.732 & 0.009 & 0.674 & 0.012 & 0.473 & 0.312 & 0.086 & 0.034 \\
21066 & vB~86     &  7.045 & 0.007 & 3.746 & 0.040 & 0.465 & 0.004 &   0.529 &   0.016 & 5.893 & 0.020 & 1.152 & 0.021 & 0.246 & 0.032 & 0.051 & 0.028 \\
21099 & vB~87     &  8.590 & 0.010 & 5.276 & 0.048 & 0.742 & 0.004 &   0.747 &   0.016 & 6.884 & 0.022 & 1.706 & 0.024 & 0.375 & 0.020 & 0.081 & 0.063 \\
21112 & vB~88     &  7.775 & 0.007 & 4.238 & 0.049 & 0.539 & 0.004 &   0.591 &   0.016 & 6.464 & 0.010 & 1.311 & 0.012 & 0.278 & 0.019 & 0.117 & 0.051 \\
21138 & vB~191    & 11.055 & 0.015 & 7.678 & 0.121 & 1.293 & 0.003 &   1.471 &   0.016 & 7.895 & 0.024 & 3.160 & 0.028 & 0.772 & 0.024 & 0.139 & 0.020 \\
21152 & vB~90     &  6.391 & 0.010 & 3.280 & 0.038 & 0.413 & 0.004 &   0.490 &   0.016 & 5.333 & 0.021 & 1.058 & 0.023 & 0.260 & 0.023 & 0.052 & 0.020 \\
21256 & L66       & 10.687 & 0.015 & 7.515 & 0.060 & 1.236 & 0.004 &   1.362 &   0.016 & 7.686 & 0.026 & 3.001 & 0.030 & 0.741 & 0.026 & 0.149 & 0.025 \\
21261 & L65       & 10.737 & 0.015 & 7.410 & 0.070 & 1.198 & 0.003 &   1.317 &   0.016 & 7.820 & 0.022 & 2.917 & 0.027 & 0.715 & 0.021 & 0.157 & 0.021 \\
21267 & vB~94     &  6.620 & 0.010 & 3.319 & 0.044 & 0.431 & 0.004 &   0.482 &   0.016 & 5.558 & 0.022 & 1.062 & 0.024 & 0.201 & 0.020 & 0.054 & 0.020 \\
21317 & vB~97     &  7.920 & 0.010 & 4.633 & 0.046 & 0.635 & 0.004 &   0.638 &   0.016 & 6.454 & 0.010 & 1.466 & 0.014 & 0.300 & 0.022 & 0.107 & 0.034 \\
21637 & vB~105    &  7.527 & 0.010 & 4.364 & 0.036 & 0.576 & 0.003 &   0.607 &   0.016 & 6.162 & 0.022 & 1.365 & 0.024 & 0.292 & 0.021 & 0.036 & 0.018 \\
21723 & L80       & 10.043 & 0.015 & 6.847 & 0.066 & 1.072 & 0.004 &   1.136 &   0.016 & 7.480 & 0.022 & 2.563 & 0.027 & 0.609 & 0.021 & 0.122 & 0.024 \\
21741 & vB~109    &  9.397 & 0.007 & 5.495 & 0.058 & 0.812 & 0.002 &   0.816 &   0.011 & 7.565 & 0.020 & 1.832 & 0.021 & 0.431 & 0.017 & 0.066 & 0.015 \\
22044 & vB~111    &  5.401 & 0.007 & 2.184 & 0.045 & 0.251 & 0.002 &   0.296 &   0.016 & 4.733 & 0.020 & 0.668 & 0.021 & 0.229 & 0.248 &--0.001 & 0.026 \\
22177 & L119      & 10.910 & 0.010 & 7.623 & 0.091 & 1.276 & 0.004 &   1.404 &   0.020 & 7.826 & 0.020 & 3.084 & 0.022 & 0.744 & 0.015 & 0.158 & 0.035 \\
22253 & L93       & 10.687 & 0.006 & 6.999 & 0.066 & 1.112 & 0.003 &   1.184 &   0.016 & 7.995 & 0.033 & 2.692 & 0.034 & 0.695 & 0.045 & 0.139 & 0.036 \\
22380 & vB~116    &  8.990 & 0.010 & 5.592 & 0.062 & 0.828 & 0.002 &   0.846 &   0.011 & 7.059 & 0.026 & 1.931 & 0.028 & 0.427 & 0.027 & 0.082 & 0.028 \\
22422 & vB~118    &  7.737 & 0.010 & 4.327 & 0.048 & 0.580 & 0.003 &   0.607 &   0.016 & 6.355 & 0.016 & 1.382 & 0.019 & 0.287 & 0.015 & 0.078 & 0.017 \\
22566 & vB~143    &  7.895 & 0.007 & 3.975 & 0.059 & 0.526 & 0.003 &   0.568 &   0.016 & 6.671 & 0.021 & 1.224 & 0.022 & 0.259 & 0.031 & 0.015 & 0.017 \\
22654 & L98       & 10.290 & 0.010 & 6.704 & 0.081 & 1.070 & 0.008 &   1.114 &   0.020 & 7.742 & 0.020 & 2.548 & 0.022 & 0.633 & 0.025 & 0.143 & 0.026 \\
22850 & vB~126    &  6.371 & 0.007 & 2.378 & 0.056 & 0.291 & 0.004 &   0.350 &   0.016 & 5.623 & 0.018 & 0.748 & 0.019 & 0.113 & 0.022 & 0.035 & 0.014 \\
23069 & vB~127    &  8.890 & 0.007 & 5.175 & 0.073 & 0.737 & 0.002 &   0.724 &   0.016 & 7.235 & 0.017 & 1.655 & 0.018 & 0.334 & 0.013 & 0.033 & 0.054 \\
23214 & vB~128    &  6.755 & 0.007 & 3.593 & 0.039 & 0.450 & 0.004 &   0.498 &   0.016 & 5.645 & 0.016 & 1.110 & 0.017 & 0.211 & 0.018 & 0.049 & 0.030 \\
23312 & BD+04:810 &  9.722 & 0.010 & 6.159 & 0.074 & 0.957 & 0.047 & \nodata & \nodata & 7.589 & 0.026 & 2.133 & 0.028 & 0.524 & 0.026 & 0.103 & 0.039 \\
23497 & vB~129    &  4.640 & 0.010 & 1.041 & 0.045 & 0.150 & 0.003 &   0.195 &   0.016 & 4.245 & 0.021 & 0.395 & 0.023 & 0.088 & 0.320 & 0.130 & 0.209 \\
23750 & BD+17:841 &  8.820 & 0.010 & 5.197 & 0.061 & 0.730 & 0.015 & \nodata & \nodata & 7.128 & 0.017 & 1.692 & 0.020 & 0.386 & 0.011 & 0.088 & 0.044 \\
24116 & BD+20:897 &  7.852 & 0.010 & 3.314 & 0.084 & 0.445 & 0.015 & \nodata & \nodata & 6.821 & 0.017 & 1.031 & 0.020 & 0.214 & 0.021 & 0.059 & 0.013 \\
24923 & BD+11:772 &  9.042 & 0.010 & 5.292 & 0.088 & 0.765 & 0.029 & \nodata & \nodata & 7.302 & 0.021 & 1.740 & 0.023 & 0.394 & 0.021 & 0.095 & 0.061 \\
26382 & vB~168    &  5.541 & 0.006 & 2.014 & 0.072 & 0.222 & 0.004 &   0.257 &   0.016 & 4.935 & 0.025 & 0.606 & 0.026 & 0.127 & 0.027 & 0.067 & 0.031 \\
28356 & BD+02:1102&  7.780 & 0.010 & 3.422 & 0.204 & 0.461 & 0.015 & \nodata & \nodata & 6.685 & 0.020 & 1.095 & 0.022 & 0.260 & 0.020 & 0.040 & 0.026 \\ 
\enddata
\tablenotetext{a}{G = Giclas.  L = Leiden (Pels).  LP = Luyten Palomar.
LTT = Luyten Two Tenths.  vB = van Buren.  vA = van Altena.}
\end{deluxetable}

\clearpage \begin{deluxetable}{llllllrrr}\footnotesize
\tablecaption{Empirically Calibrated Hyades Isochrone}
\tablewidth{0pt}
\tabletypesize{\footnotesize}
\tablehead{
 \colhead{$T_{\rm eff}$(K)} &
 \colhead{$M / M_\odot$} &
 \colhead{$M_V$} & 
 \colhead{$B - V$} &
 \colhead{$V - I_C$} &
 \colhead{$V - K_s$} &
 \colhead{$\Delta(B - V)$} &
 \colhead{$\Delta(V - I_C$)} &
 \colhead{$\Delta(V - K_s$)}
}
\startdata
8280	&	2.235	&	1.10	&	0.139	&	0.162	&	0.374	&	0.027	&	0.053	&	0.014	\\
8276	&	2.214	&	1.15	&	0.141	&	0.164	&	0.377	&	0.027	&	0.054	&	0.015	\\
8270	&	2.192	&	1.20	&	0.143	&	0.167	&	0.380	&	0.027	&	0.054	&	0.016	\\
8262	&	2.170	&	1.25	&	0.146	&	0.170	&	0.384	&	0.027	&	0.055	&	0.017	\\
8251	&	2.149	&	1.30	&	0.149	&	0.174	&	0.390	&	0.027	&	0.056	&	0.018	\\
8236	&	2.127	&	1.35	&	0.153	&	0.178	&	0.396	&	0.027	&	0.057	&	0.019	\\
8215	&	2.105	&	1.40	&	0.158	&	0.184	&	0.409	&	0.027	&	0.057	&	0.024	\\
8189	&	2.083	&	1.45	&	0.163	&	0.191	&	0.425	&	0.027	&	0.058	&	0.032	\\
8158	&	2.062	&	1.50	&	0.170	&	0.199	&	0.443	&	0.026	&	0.058	&	0.040	\\
8124	&	2.040	&	1.55	&	0.176	&	0.208	&	0.461	&	0.026	&	0.059	&	0.046	\\
8088	&	2.019	&	1.60	&	0.183	&	0.218	&	0.479	&	0.025	&	0.059	&	0.051	\\
8052	&	1.997	&	1.65	&	0.190	&	0.228	&	0.496	&	0.024	&	0.059	&	0.056	\\
8015	&	1.976	&	1.70	&	0.197	&	0.238	&	0.514	&	0.023	&	0.059	&	0.061	\\
7978	&	1.955	&	1.75	&	0.204	&	0.247	&	0.532	&	0.021	&	0.058	&	0.065	\\
7942	&	1.934	&	1.80	&	0.210	&	0.256	&	0.549	&	0.020	&	0.057	&	0.069	\\
7905	&	1.914	&	1.85	&	0.216	&	0.265	&	0.566	&	0.018	&	0.055	&	0.072	\\
7868	&	1.893	&	1.90	&	0.223	&	0.273	&	0.582	&	0.016	&	0.053	&	0.075	\\
7830	&	1.873	&	1.95	&	0.229	&	0.281	&	0.599	&	0.014	&	0.051	&	0.077	\\
7790	&	1.854	&	2.00	&	0.235	&	0.287	&	0.614	&	0.012	&	0.049	&	0.079	\\
7749	&	1.834	&	2.05	&	0.242	&	0.294	&	0.630	&	0.010	&	0.048	&	0.081	\\
7706	&	1.815	&	2.10	&	0.249	&	0.300	&	0.646	&	0.008	&	0.046	&	0.082	\\
7662	&	1.796	&	2.15	&	0.256	&	0.305	&	0.662	&	0.006	&	0.045	&	0.083	\\
7616	&	1.776	&	2.20	&	0.263	&	0.310	&	0.678	&	0.004	&	0.045	&	0.084	\\
7568	&	1.757	&	2.25	&	0.271	&	0.315	&	0.694	&	0.002	&	0.044	&	0.085	\\
7520	&	1.738	&	2.30	&	0.279	&	0.320	&	0.709	&	0.001	&	0.044	&	0.085	\\
7471	&	1.720	&	2.35	&	0.286	&	0.324	&	0.724	&	--0.001	&	0.043	&	0.085	\\
7423	&	1.702	&	2.40	&	0.294	&	0.329	&	0.739	&	--0.002	&	0.043	&	0.085	\\
7375	&	1.685	&	2.45	&	0.301	&	0.334	&	0.754	&	--0.002	&	0.042	&	0.085	\\
7327	&	1.667	&	2.50	&	0.308	&	0.341	&	0.769	&	--0.003	&	0.040	&	0.084	\\
7280	&	1.651	&	2.55	&	0.315	&	0.348	&	0.784	&	--0.003	&	0.039	&	0.084	\\
7232	&	1.634	&	2.60	&	0.322	&	0.356	&	0.799	&	--0.003	&	0.038	&	0.083	\\
7186	&	1.618	&	2.65	&	0.329	&	0.365	&	0.814	&	--0.004	&	0.037	&	0.083	\\
7139	&	1.603	&	2.70	&	0.336	&	0.375	&	0.829	&	--0.004	&	0.036	&	0.083	\\
7092	&	1.587	&	2.75	&	0.343	&	0.386	&	0.845	&	--0.004	&	0.035	&	0.082	\\
7046	&	1.571	&	2.80	&	0.350	&	0.397	&	0.860	&	--0.004	&	0.035	&	0.081	\\
7001	&	1.556	&	2.85	&	0.358	&	0.408	&	0.876	&	--0.004	&	0.034	&	0.080	\\
6956	&	1.541	&	2.90	&	0.365	&	0.419	&	0.892	&	--0.005	&	0.033	&	0.079	\\
6913	&	1.526	&	2.95	&	0.372	&	0.429	&	0.909	&	--0.005	&	0.032	&	0.077	\\
6871	&	1.511	&	3.00	&	0.379	&	0.439	&	0.926	&	--0.006	&	0.031	&	0.075	\\
6830	&	1.496	&	3.05	&	0.387	&	0.449	&	0.943	&	--0.006	&	0.031	&	0.072	\\
6792	&	1.482	&	3.10	&	0.394	&	0.457	&	0.961	&	--0.007	&	0.029	&	0.069	\\
6754	&	1.468	&	3.15	&	0.401	&	0.465	&	0.980	&	--0.008	&	0.028	&	0.065	\\
6718	&	1.454	&	3.20	&	0.408	&	0.472	&	0.998	&	--0.009	&	0.027	&	0.062	\\
6684	&	1.440	&	3.25	&	0.415	&	0.479	&	1.016	&	--0.010	&	0.025	&	0.057	\\
6650	&	1.426	&	3.30	&	0.422	&	0.485	&	1.034	&	--0.010	&	0.023	&	0.052	\\
6618	&	1.413	&	3.35	&	0.429	&	0.491	&	1.051	&	--0.011	&	0.021	&	0.047	\\
6586	&	1.400	&	3.40	&	0.436	&	0.497	&	1.068	&	--0.012	&	0.020	&	0.041	\\
6556	&	1.387	&	3.45	&	0.443	&	0.503	&	1.084	&	--0.012	&	0.018	&	0.034	\\
6525	&	1.375	&	3.50	&	0.450	&	0.508	&	1.099	&	--0.012	&	0.016	&	0.028	\\
6496	&	1.362	&	3.55	&	0.457	&	0.514	&	1.116	&	--0.012	&	0.015	&	0.023	\\
6466	&	1.350	&	3.60	&	0.464	&	0.521	&	1.132	&	--0.012	&	0.014	&	0.017	\\
6438	&	1.338	&	3.65	&	0.471	&	0.527	&	1.148	&	--0.012	&	0.013	&	0.012	\\
6409	&	1.326	&	3.70	&	0.478	&	0.534	&	1.164	&	--0.011	&	0.012	&	0.008	\\
6382	&	1.315	&	3.75	&	0.485	&	0.540	&	1.181	&	--0.010	&	0.011	&	0.003	\\
6354	&	1.303	&	3.80	&	0.493	&	0.546	&	1.197	&	--0.010	&	0.010	&	0.000	\\
6327	&	1.292	&	3.85	&	0.500	&	0.552	&	1.214	&	--0.009	&	0.010	&	--0.003	\\
6300	&	1.281	&	3.90	&	0.507	&	0.558	&	1.231	&	--0.007	&	0.008	&	--0.007	\\
6273	&	1.269	&	3.95	&	0.515	&	0.564	&	1.247	&	--0.006	&	0.007	&	--0.010	\\
6246	&	1.259	&	4.00	&	0.522	&	0.570	&	1.264	&	--0.005	&	0.006	&	--0.013	\\
6220	&	1.248	&	4.05	&	0.530	&	0.576	&	1.281	&	--0.003	&	0.005	&	--0.015	\\
6194	&	1.237	&	4.10	&	0.538	&	0.582	&	1.299	&	--0.002	&	0.004	&	--0.017	\\
6167	&	1.227	&	4.15	&	0.545	&	0.589	&	1.318	&	0.000	&	0.003	&	--0.018	\\
6141	&	1.217	&	4.20	&	0.553	&	0.595	&	1.337	&	0.002	&	0.002	&	--0.019	\\
6115	&	1.207	&	4.25	&	0.561	&	0.601	&	1.357	&	0.003	&	0.001	&	--0.018	\\
6089	&	1.197	&	4.30	&	0.569	&	0.608	&	1.377	&	0.005	&	0.000	&	--0.018	\\
6063	&	1.187	&	4.35	&	0.577	&	0.614	&	1.396	&	0.007	&	0.000	&	--0.018	\\
6036	&	1.177	&	4.40	&	0.585	&	0.621	&	1.414	&	0.009	&	--0.001	&	--0.018	\\
6010	&	1.167	&	4.45	&	0.594	&	0.628	&	1.430	&	0.010	&	--0.001	&	--0.019	\\
5982	&	1.157	&	4.50	&	0.602	&	0.636	&	1.445	&	0.012	&	--0.002	&	--0.020	\\
5955	&	1.147	&	4.55	&	0.611	&	0.643	&	1.460	&	0.013	&	--0.002	&	--0.021	\\
5927	&	1.137	&	4.60	&	0.619	&	0.651	&	1.474	&	0.015	&	--0.003	&	--0.021	\\
5899	&	1.128	&	4.65	&	0.628	&	0.658	&	1.488	&	0.016	&	--0.003	&	--0.021	\\
5871	&	1.118	&	4.70	&	0.637	&	0.666	&	1.503	&	0.018	&	--0.004	&	--0.020	\\
5843	&	1.108	&	4.75	&	0.646	&	0.674	&	1.519	&	0.019	&	--0.005	&	--0.019	\\
5815	&	1.099	&	4.80	&	0.654	&	0.682	&	1.535	&	0.020	&	--0.005	&	--0.018	\\
5787	&	1.090	&	4.85	&	0.663	&	0.690	&	1.551	&	0.021	&	--0.006	&	--0.018	\\
5760	&	1.080	&	4.90	&	0.672	&	0.698	&	1.568	&	0.021	&	--0.006	&	--0.018	\\
5732	&	1.071	&	4.95	&	0.681	&	0.706	&	1.586	&	0.022	&	--0.007	&	--0.018	\\
5704	&	1.062	&	5.00	&	0.690	&	0.715	&	1.605	&	0.022	&	--0.007	&	--0.019	\\
5676	&	1.054	&	5.05	&	0.700	&	0.724	&	1.625	&	0.023	&	--0.007	&	--0.020	\\
5648	&	1.045	&	5.10	&	0.709	&	0.733	&	1.646	&	0.023	&	--0.007	&	--0.021	\\
5620	&	1.036	&	5.15	&	0.719	&	0.742	&	1.667	&	0.023	&	--0.007	&	--0.024	\\
5591	&	1.028	&	5.20	&	0.729	&	0.751	&	1.689	&	0.023	&	--0.008	&	--0.026	\\
5561	&	1.020	&	5.25	&	0.739	&	0.760	&	1.711	&	0.023	&	--0.008	&	--0.029	\\
5532	&	1.012	&	5.30	&	0.749	&	0.768	&	1.733	&	0.023	&	--0.008	&	--0.032	\\
5502	&	1.003	&	5.35	&	0.759	&	0.777	&	1.756	&	0.023	&	--0.008	&	--0.036	\\
5471	&	0.995	&	5.40	&	0.769	&	0.786	&	1.779	&	0.023	&	--0.008	&	--0.040	\\
5440	&	0.988	&	5.45	&	0.780	&	0.795	&	1.802	&	0.023	&	--0.008	&	--0.044	\\
5408	&	0.980	&	5.50	&	0.790	&	0.804	&	1.824	&	0.022	&	--0.007	&	--0.048	\\
5377	&	0.972	&	5.55	&	0.801	&	0.813	&	1.847	&	0.021	&	--0.006	&	--0.052	\\
5344	&	0.964	&	5.60	&	0.811	&	0.823	&	1.870	&	0.020	&	--0.004	&	--0.056	\\
5312	&	0.957	&	5.65	&	0.822	&	0.833	&	1.893	&	0.019	&	--0.002	&	--0.059	\\
5279	&	0.949	&	5.70	&	0.833	&	0.843	&	1.917	&	0.017	&	0.001	&	--0.061	\\
5245	&	0.942	&	5.75	&	0.844	&	0.854	&	1.941	&	0.015	&	0.004	&	--0.061	\\
5212	&	0.934	&	5.80	&	0.854	&	0.866	&	1.966	&	0.013	&	0.008	&	--0.060	\\
5179	&	0.927	&	5.85	&	0.865	&	0.877	&	1.993	&	0.011	&	0.011	&	--0.058	\\
5146	&	0.920	&	5.90	&	0.876	&	0.888	&	2.019	&	0.008	&	0.015	&	--0.056	\\
5113	&	0.913	&	5.95	&	0.887	&	0.900	&	2.046	&	0.005	&	0.018	&	--0.054	\\
5080	&	0.906	&	6.00	&	0.898	&	0.911	&	2.072	&	0.002	&	0.020	&	--0.053	\\
5048	&	0.899	&	6.05	&	0.908	&	0.921	&	2.098	&	--0.001	&	0.023	&	--0.052	\\
5017	&	0.892	&	6.10	&	0.919	&	0.933	&	2.125	&	--0.004	&	0.025	&	--0.053	\\
4986	&	0.886	&	6.15	&	0.930	&	0.944	&	2.151	&	--0.006	&	0.027	&	--0.053	\\
4956	&	0.879	&	6.20	&	0.941	&	0.955	&	2.179	&	--0.009	&	0.029	&	--0.054	\\
4927	&	0.873	&	6.25	&	0.952	&	0.966	&	2.206	&	--0.012	&	0.031	&	--0.056	\\
4898	&	0.867	&	6.30	&	0.963	&	0.978	&	2.235	&	--0.014	&	0.032	&	--0.058	\\
4870	&	0.861	&	6.35	&	0.974	&	0.990	&	2.263	&	--0.016	&	0.033	&	--0.060	\\
4843	&	0.855	&	6.40	&	0.985	&	1.003	&	2.293	&	--0.019	&	0.034	&	--0.061	\\
4816	&	0.850	&	6.45	&	0.996	&	1.016	&	2.322	&	--0.021	&	0.034	&	--0.063	\\
4790	&	0.844	&	6.50	&	1.007	&	1.030	&	2.353	&	--0.022	&	0.035	&	--0.064	\\
4765	&	0.839	&	6.55	&	1.019	&	1.044	&	2.384	&	--0.024	&	0.035	&	--0.065	\\
4740	&	0.833	&	6.60	&	1.030	&	1.059	&	2.415	&	--0.026	&	0.034	&	--0.065	\\
4715	&	0.828	&	6.65	&	1.041	&	1.074	&	2.446	&	--0.027	&	0.034	&	--0.065	\\
4691	&	0.823	&	6.70	&	1.053	&	1.090	&	2.478	&	--0.028	&	0.033	&	--0.065	\\
4667	&	0.818	&	6.75	&	1.064	&	1.106	&	2.510	&	--0.029	&	0.032	&	--0.064	\\
4644	&	0.813	&	6.80	&	1.076	&	1.122	&	2.542	&	--0.030	&	0.031	&	--0.062	\\
4620	&	0.808	&	6.85	&	1.087	&	1.138	&	2.574	&	--0.031	&	0.030	&	--0.060	\\
4597	&	0.803	&	6.90	&	1.099	&	1.155	&	2.606	&	--0.031	&	0.028	&	--0.057	\\
4573	&	0.798	&	6.95	&	1.110	&	1.171	&	2.638	&	--0.031	&	0.027	&	--0.054	\\
4550	&	0.793	&	7.00	&	1.121	&	1.187	&	2.670	&	--0.031	&	0.025	&	--0.050	\\
4527	&	0.788	&	7.05	&	1.132	&	1.203	&	2.702	&	--0.031	&	0.024	&	--0.045	\\
4503	&	0.783	&	7.10	&	1.144	&	1.219	&	2.734	&	--0.030	&	0.023	&	--0.039	\\
4480	&	0.778	&	7.15	&	1.155	&	1.235	&	2.765	&	--0.029	&	0.021	&	--0.033	\\
4457	&	0.773	&	7.20	&	1.167	&	1.252	&	2.798	&	--0.027	&	0.021	&	--0.026	\\
4435	&	0.768	&	7.25	&	1.178	&	1.268	&	2.830	&	--0.025	&	0.021	&	--0.018	\\
4412	&	0.763	&	7.30	&	1.190	&	1.285	&	2.862	&	--0.022	&	0.021	&	--0.009	\\
4390	&	0.757	&	7.35	&	1.202	&	1.302	&	2.894	&	--0.019	&	0.022	&	0.000	\\
4368	&	0.752	&	7.40	&	1.214	&	1.320	&	2.927	&	--0.015	&	0.023	&	0.011	\\
4347	&	0.747	&	7.45	&	1.226	&	1.339	&	2.961	&	--0.010	&	0.026	&	0.023	\\
4326	&	0.742	&	7.50	&	1.238	&	1.357	&	2.994	&	--0.006	&	0.029	&	0.035	\\
4305	&	0.737	&	7.55	&	1.250	&	1.378	&	3.028	&	--0.001	&	0.034	&	0.048	\\
4285	&	0.731	&	7.60	&	1.262	&	1.400	&	3.061	&	0.004	&	0.039	&	0.061	\\
4265	&	0.726	&	7.65	&	1.274	&	1.422	&	3.093	&	0.009	&	0.045	&	0.073	\\
4246	&	0.721	&	7.70	&	1.285	&	1.445	&	3.123	&	0.014	&	0.052	&	0.083	\\
4228	&	0.715	&	7.75	&	1.296	&	1.468	&	3.151	&	0.018	&	0.059	&	0.092	\\
4209	&	0.710	&	7.80	&	1.307	&	1.493	&	3.177	&	0.023	&	0.066	&	0.098	\\
4192	&	0.705	&	7.85	&	1.317	&	1.517	&	3.199	&	0.028	&	0.074	&	0.102	\\
4175	&	0.700	&	7.90	&	1.328	&	1.541	&	3.221	&	0.032	&	0.080	&	0.104	\\
4158	&	0.694	&	7.95	&	1.338	&	1.562	&	3.244	&	0.037	&	0.084	&	0.108	\\
4142	&	0.689	&	8.00	&	1.348	&	1.580	&	3.270	&	0.042	&	0.085	&	0.115	\\
4126	&	0.684	&	8.05	&	1.358	&	1.596	&	3.303	&	0.046	&	0.085	&	0.129	\\
4110	&	0.678	&	8.10	&	1.368	&	1.616	&	3.341	&	0.051	&	0.087	&	0.148	\\
4095	&	0.673	&	8.15	&	1.377	&	1.643	&	3.385	&	0.055	&	0.098	&	0.172	\\
4080	&	0.668	&	8.20	&	1.386	&	1.676	&	3.431	&	0.059	&	0.115	&	0.200	\\
4066	&	0.663	&	8.25	&	1.395	&	1.711	&	3.477	&	0.063	&	0.135	&	0.226	\\
4051	&	0.657	&	8.30	&	1.403	&	1.756	&	3.508	&	0.067	&	0.165	&	0.238	\\
4038	&	0.652	&	8.35	&	1.411	&	1.801	&	3.532	&	0.071	&	0.195	&	0.242	\\
4024	&	0.647	&	8.40	&	1.419	&	1.844	&	3.554	&	0.074	&	0.224	&	0.244	\\
4010	&	0.642	&	8.45	&	1.426	&	1.882	&	3.586	&	0.077	&	0.250	&	0.256	\\
3997	&	0.636	&	8.50	&	1.432	&	1.927	&	3.624	&	0.080	&	0.282	&	0.274	\\
\enddata
\end{deluxetable}

\clearpage \begin{deluxetable}{lccccc}
\tablewidth{0pt}
\tablecaption{Metallicity sensitivity of isochrone}
\tablehead{
  \colhead{Color calibration} &
  \colhead{$B - V$} &
  \colhead{$V - I_C$} &
  \colhead{$V - K_s$} &
  \colhead{${\rm [M/H]}$} &
  \colhead{$\Delta(m-M)_V$}
}
\startdata
Lejeune et al. (1998)      & 1.402 & 0.927 & 0.804 &  $+0.142$ &   0.00  \\
Uncorrected Lejeune        & 1.432 & 0.903 & 0.767 &  $-0.019$ & $-0.10$ \\
Alonso et al. (1995,1996)  & 1.421 & 0.727 & 0.758 &  $+0.036$ & $-0.40$ \\
\enddata
\end{deluxetable}

\clearpage \figcaption[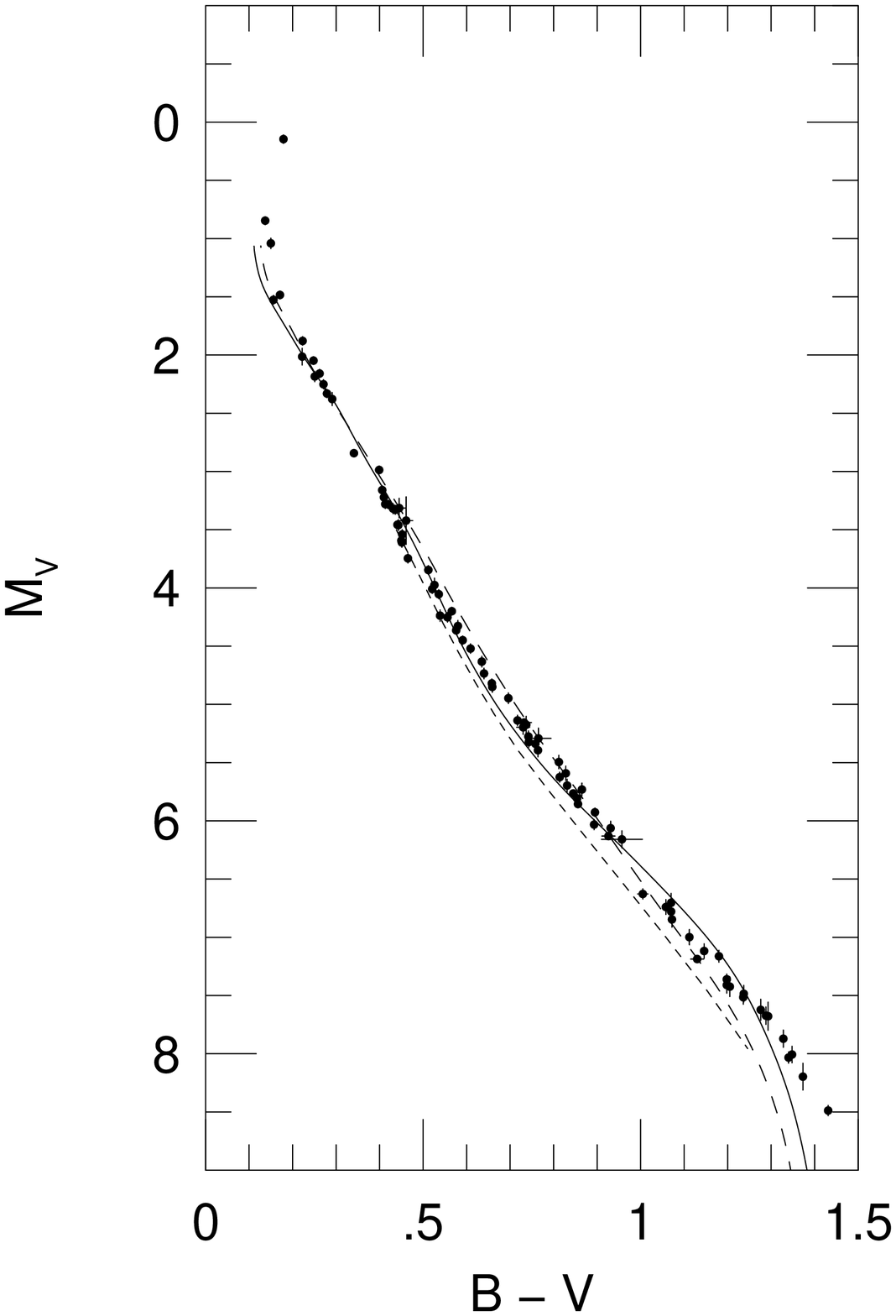]{Comparison of Hyades
single-star photometry (points) in $M_V$, $B - V$ to a
theoretical isochrone (solid line) for [Fe/H] $=+0.13$,
age 550 Myr, which uses the \citet{lejeune98} color
calibration.  Shown as a long-dashed line is the same
isochrone, but employing the uncalibrated \citet{lejeune98}
color-temperature relation; the short-dashed line is the
equivalent for the \citet{alonso95, alonso96} calibration.
Error bars are shown for all points, even though these
are typically smaller than the points themselves.}

\figcaption[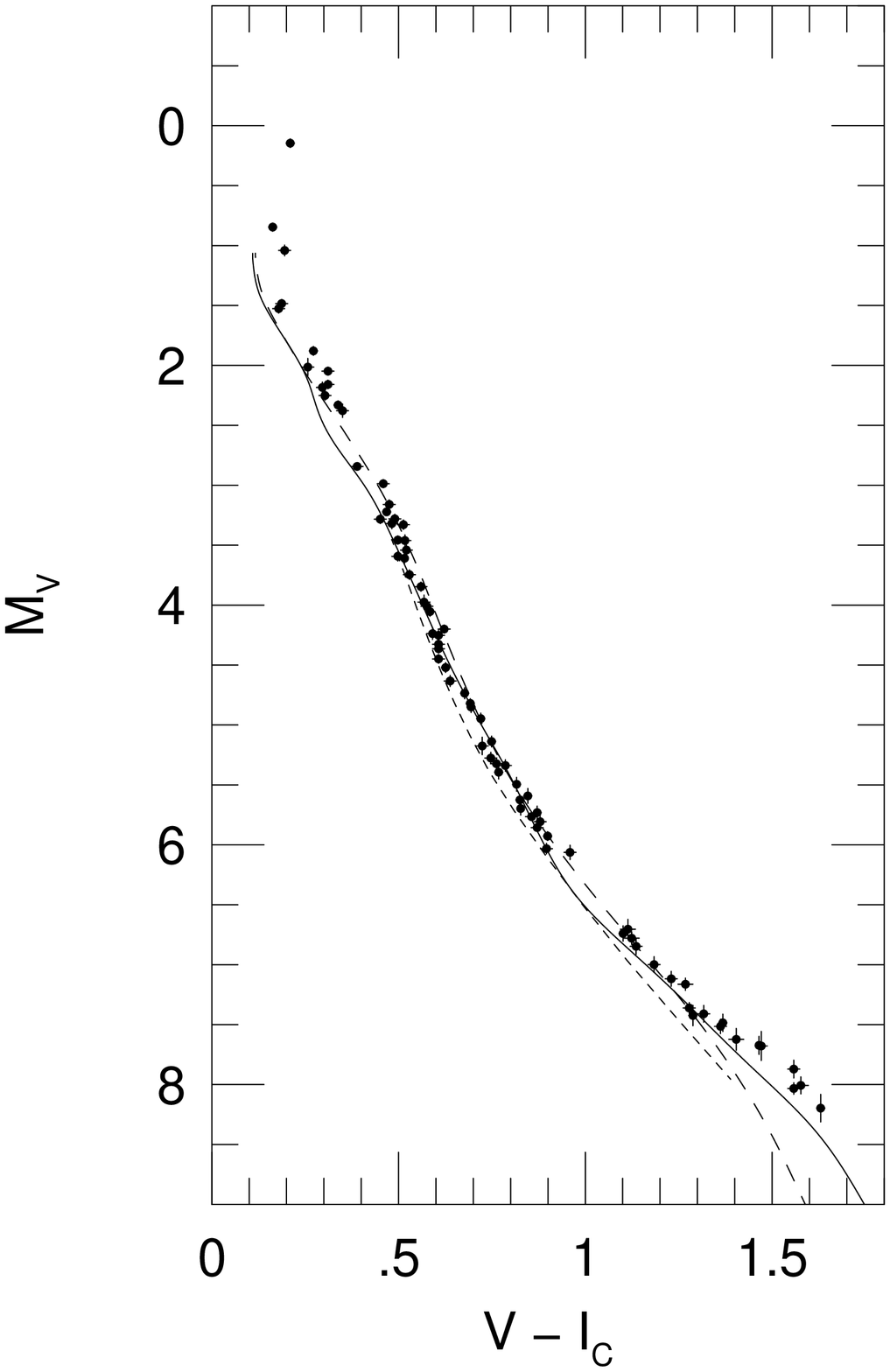]{Same as Figure 1, but in $M_V$, $V
- I_C$.}

\figcaption[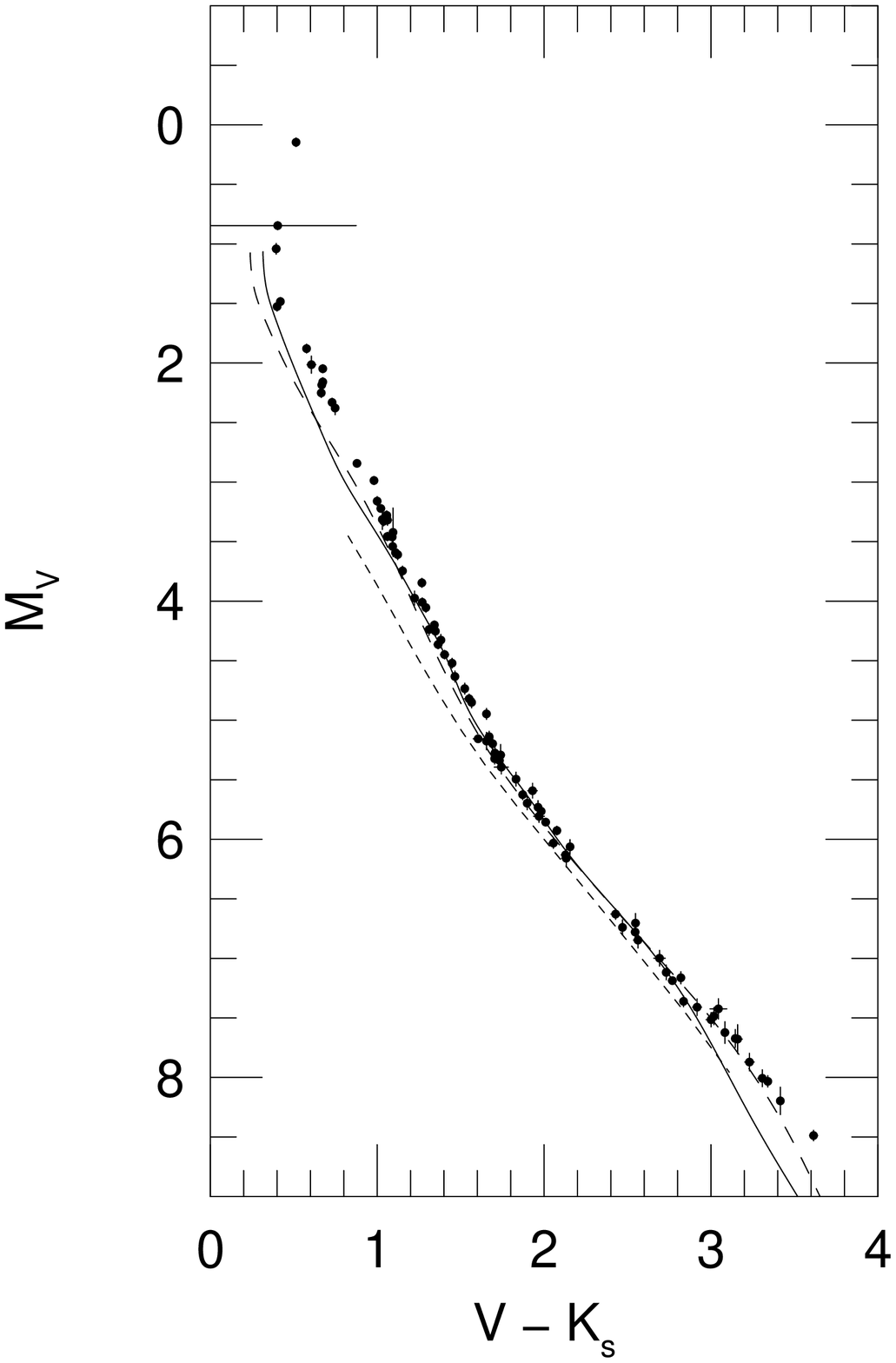]{Same as Figure 1, but in $M_V$, $V
- K_s$.}

\figcaption[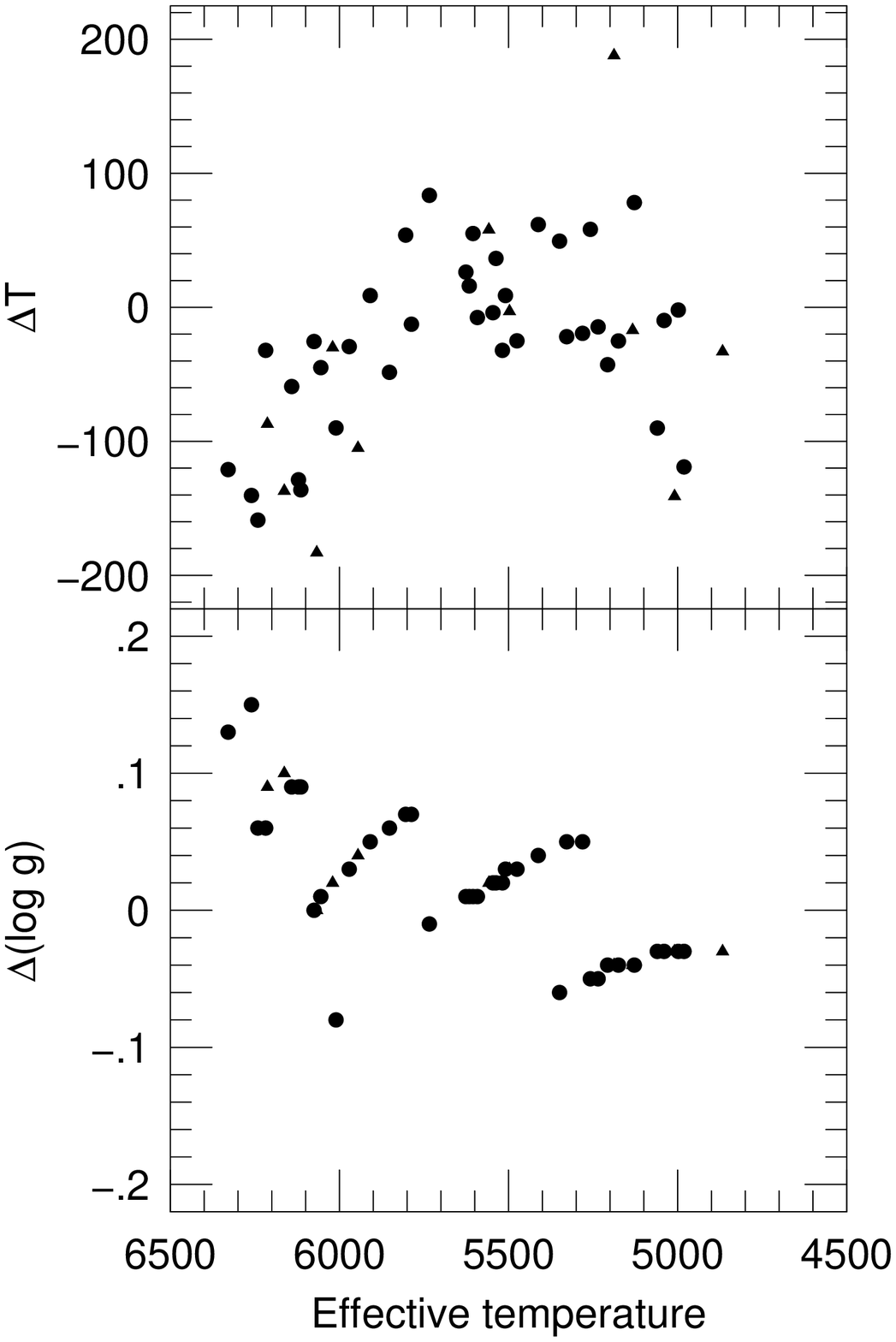]{Differences in effective temperature
(top panel) and gravity (lower panel), in the sense of
(ischrone -- spectra), between spectroscopically determined
parameters (Paulson, Sneden, \& Cochran 2003) and the
Hyades isochrone at fixed $M_V$.  While the agreement
is good throughout most of the temperature range, the
spectroscopic solution yields hotter temperatures and lower
gravities than the isochrone.  Filled circles are for the
stars used to calibrate the empirical color corrections,
while the triangles show other stars in the Paulson et al.\
sample not included in our Table 1.}

\figcaption[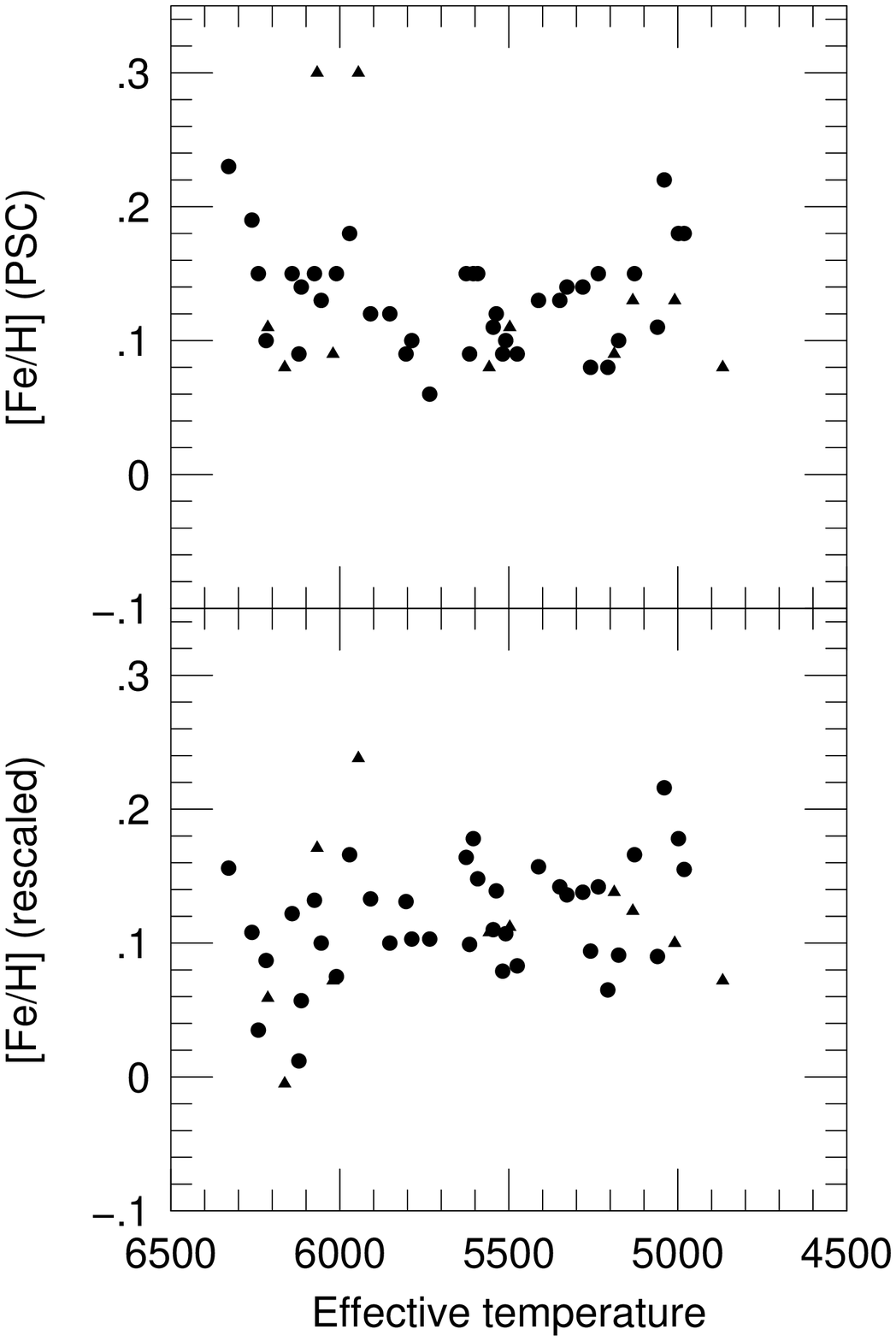]{Upper panel: iron abundances for
Hyades stars from Paulson, Sneden, \& Cochran (2003) as a
function of isochrone effective temperature.  Lower panel:
rescaled iron abundances using the isochrone temperatures
and gravities, as described in the text.  Symbols are the
same as in Figure 4.}

\figcaption[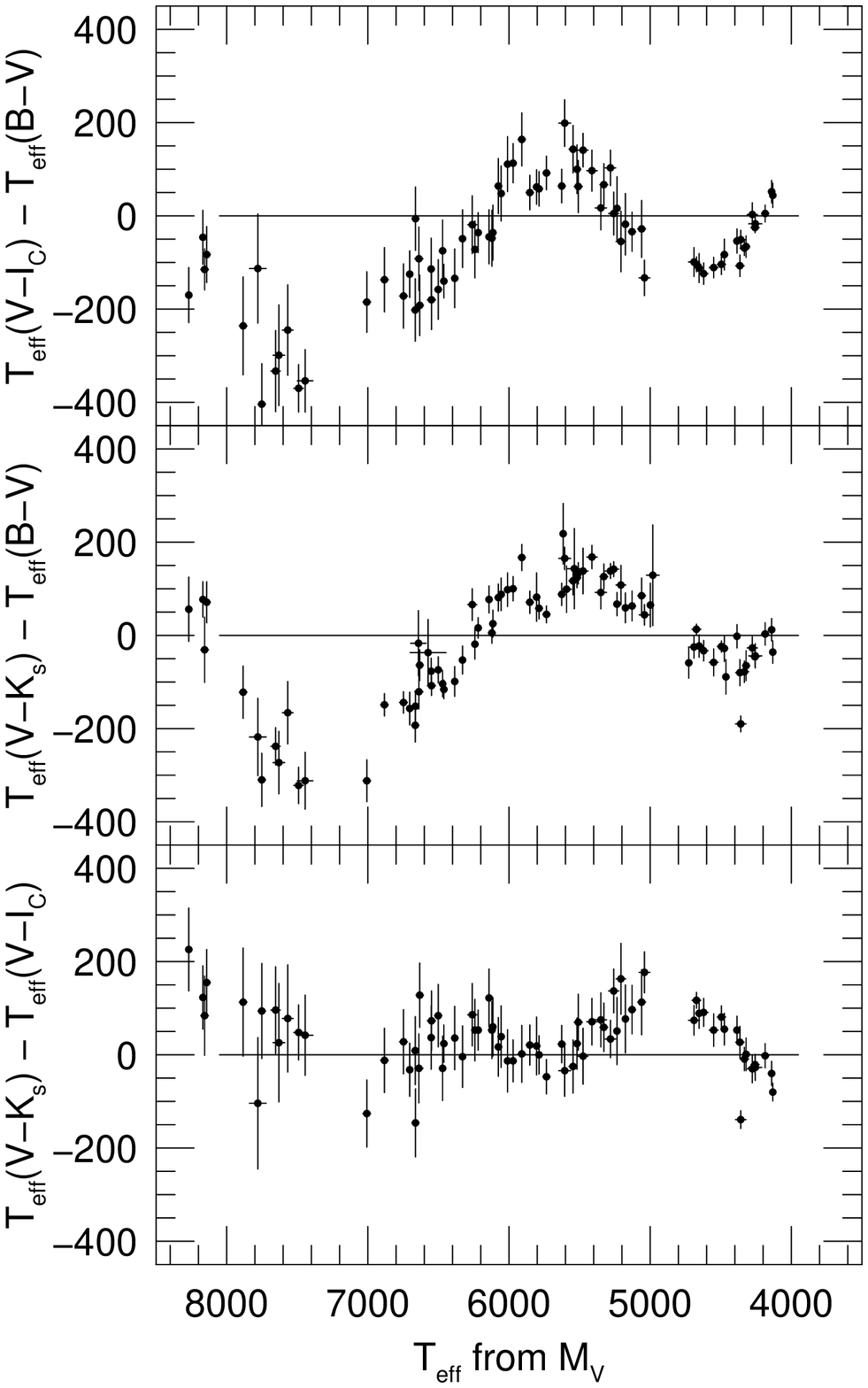]{Differences between $T_{\rm eff}$
estimates from $B-V$ and $V-I_C$ (top panel), from $B-V$
and $V-K_s$ (center panel), and from $V-I_C$ and $V-K_s$
(lower panel) for the Lejeune color calibration as a
function of $T_{\rm eff}$ estimated from the isochrone
$M_V$.}

\figcaption[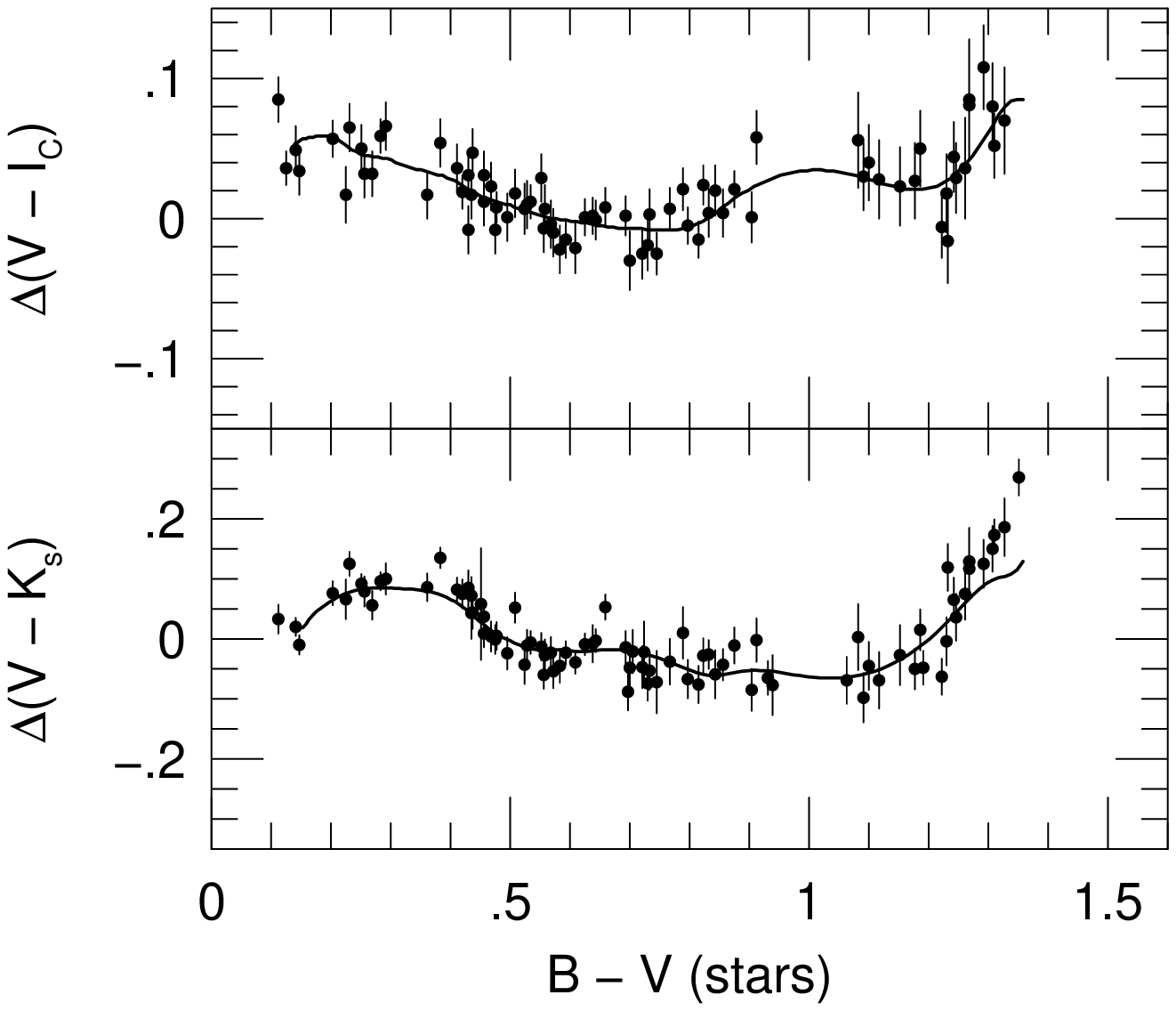]{Method of constructing color
corrections to the theoretical isochrone.  The points in
the top panel show the difference in $V - I_C$ between
the isochrone (Figure 1) and the data at the isochrone
color.  The sense of the difference is (star -- ischrone).
The smoothed line was constructed by computing a weighted
average difference in a moving window containing 3--5
stars.  The lower panels is the equivalent for $V - K_s$.
As discussed in the text, the corrections are applied to
the isochrone colors holding $M_V$ fixed.}

\figcaption[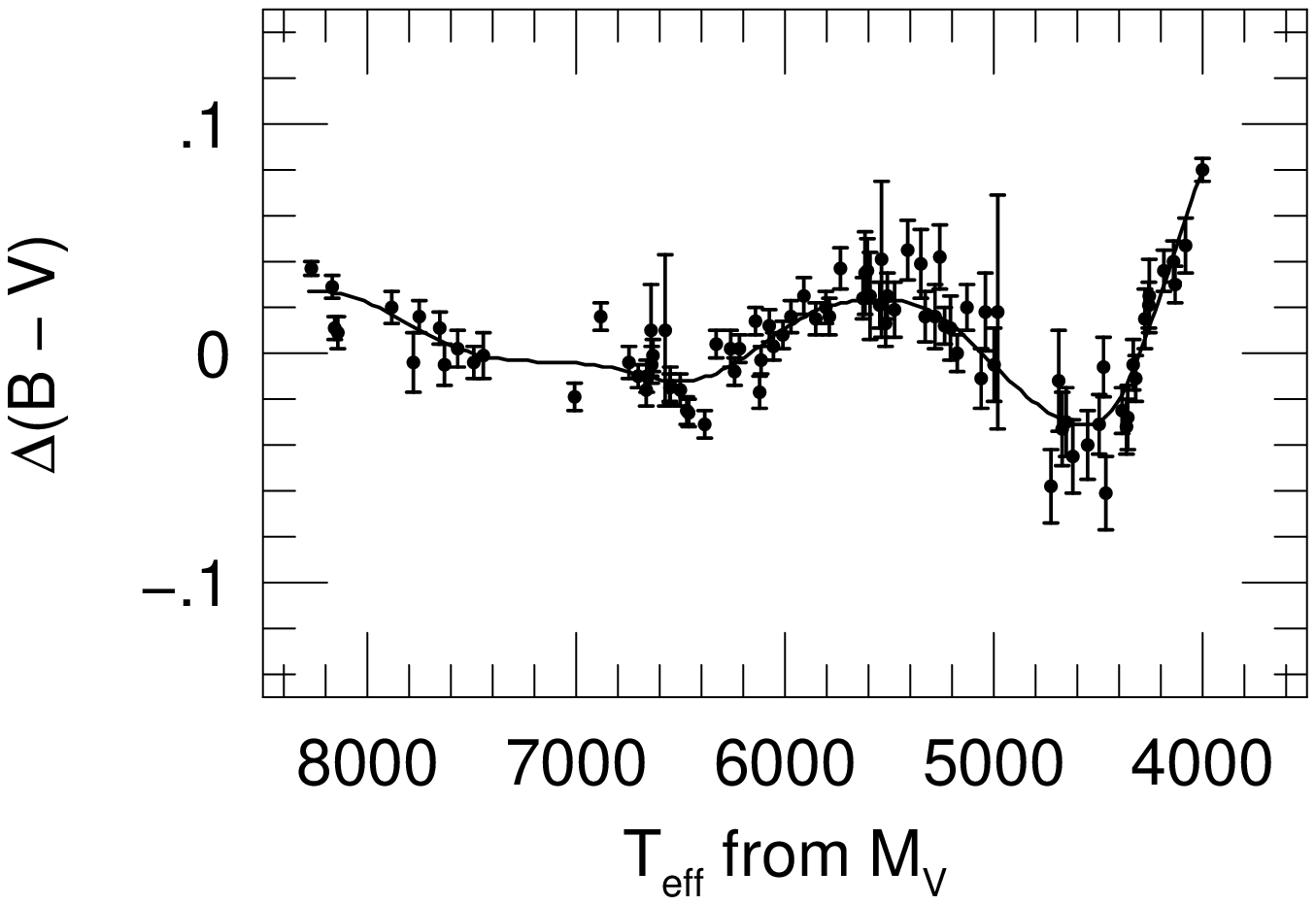]{Color correction in $B - V$.}

\figcaption[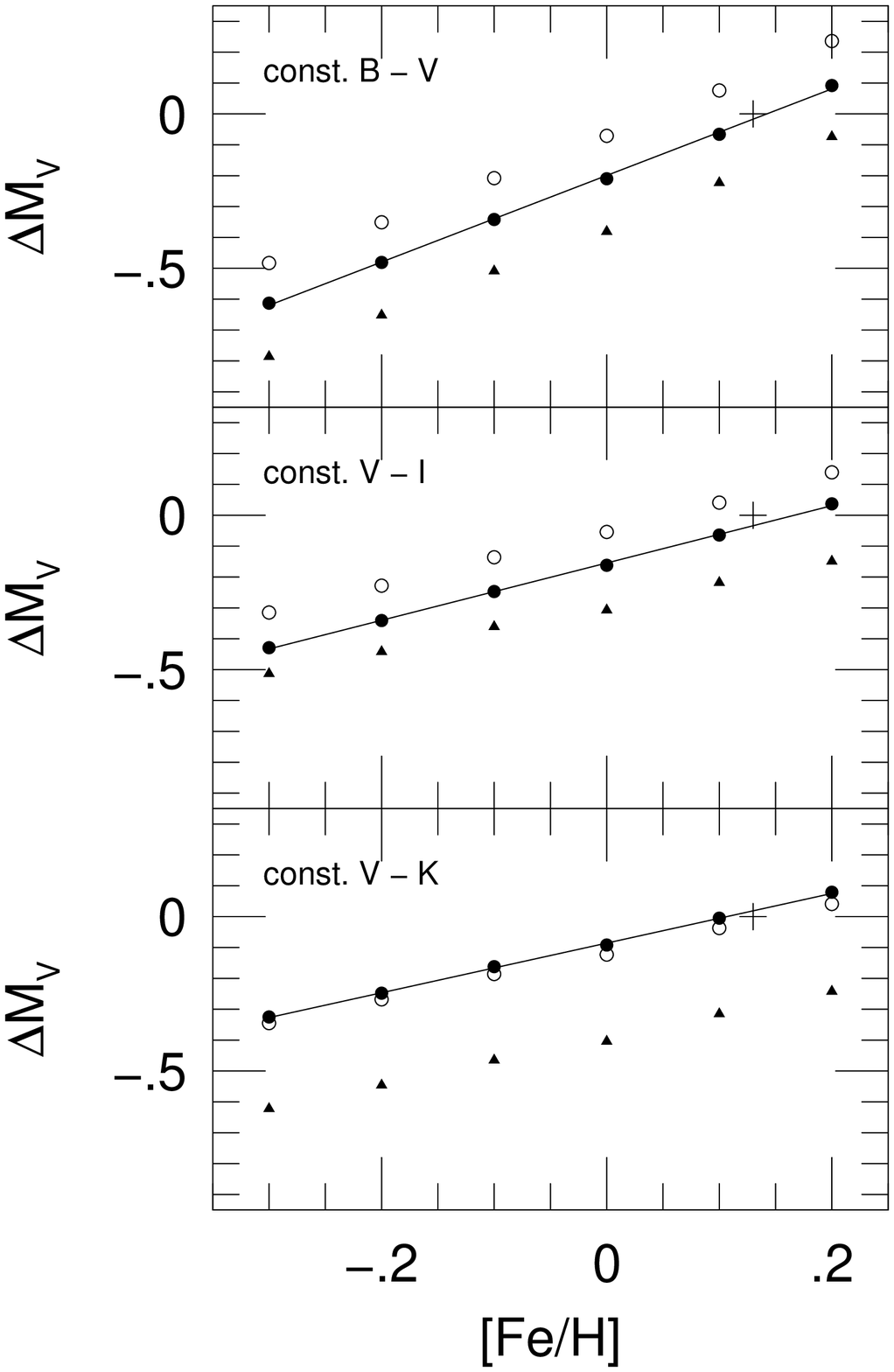]{Sensitivity of the isochrone luminosity
to metallicity, for the several color calibrations
discussed in the text.}


\begin{thebibliography}{}


\bibitem[Allard \& Hauschildt(1995)]{ah95} Allard, F., \&
Hauschildt P. H. 1995, \apj, 445, 433

\bibitem[Allende Prieto \& Lambert(1999)]{apl99} Allende
Prieto, C., \& Lambert, D. W. 1999, \aap, 352, 555

\bibitem[Alonso et al.(1995)]{alonso95} Alsonso, A.,
Arribas, S., \& Mart\'{\i}nez-Roger, C. 1995, \aap, 297, 197

\bibitem[Alonso et al.(1996)]{alonso96} Alsonso, A.,
Arribas, S., \& Mart\'{\i}nez-Roger, C. 1996, \aaps, 117, 227

\bibitem[Bahcall, Pinsonneault, \& Basu(2001)]{bpb01}
Bahcall, J. N., Pinsonneault, M. H., \& Basu, S. 2001,
\apj, 555, 990

\bibitem[Basu, Pinsonneault, \& Bahcall(2000)]{bpb00}
Basu, S., Pinsonneault, M. H., \& Bahcall, J. N. 2000,
\apj, 529, 1084

\bibitem[Belikov et al.(1998)]{bel98} Belikov, A. N.,
Hirte, S., Meusinger, H., Piskunov, A. E., \& Schilbach,
E. 1998, \aap, 332, 575

\bibitem[Bessel(1979)]{bessel79} Bessel, M. S. 1979, \pasp,
91, 589

\bibitem[Bessel \& Weis(1987)]{bw87} Bessel, M. S., \&
Weis, E. W. 1987, \pasp, 99, 642

\bibitem[Boesgaard \& Friel(1990)]{bf90} Boesgaard, A. M.,
\& Friel, E. D. 1990, \apj, 351, 467

\bibitem[de Bruijne, Hoogerwerf, \& de Zeeuw(2001)]{bhz01}
de Bruijne, J. H. J., Hoogerwerf, R., \& de Zeeuw,
P. T. 2001, \aap, 367, 111

\bibitem[Carney(1982)]{car82} Carney, B. W. 1982, \aj,
87, 1527

\bibitem[Carney \& Aaronson(1979)]{ca79} Carney B. W., \&
Aaronson M. 1979, \aj, 84, 867

\bibitem[Carpenter(2001)]{car01} Carpenter, J. M. 2001,
\aj, 121, 2851

\bibitem[Castellani, Degl'Innocenti, \& Prada Moroni(2001)]
{cas01} Castellani, V., Degl'Innocenti, S., \& Prada
Moroni, P. G. 2001, \mnras, 320, 66

\bibitem[Crawford(1975)]{cra75} Crawford, D. L. 1975, \aj,
80, 955

\bibitem[Eggen(1948)]{eggen48} Eggen, O. J. 1948, \aj,
54, 35

\bibitem[Eggen(1975)]{eggen75} Eggen, O. J. 1975, \pasp,
87, 107

\bibitem[Eggen(1982)]{eggen82} Eggen O. J., 1982, \apjs,
50, 221

\bibitem[Elias et al.(1982)]{elias82} Elias, J. H., Frogel,
J. A., Matthews, K., \& Neugebauer, G. 1982, \aj, 87, 1029

\bibitem[ESA(1997)]{esa97} European Space Agency, 1997, The
Hipparcos and Tycho Catalogues (ESA SP-1200) (Paris: ESA)

\bibitem[Flower(1996)]{flower96} Flower, P. J. 1996, \apj,
469, 355

\bibitem[Frogel et al.(1978)]{fpam78} Frogel, J. A.,
Persson, S. E., Aaronson, M., \& Matthews, K. 1978, \apj,
220, 75

\bibitem[Gatewood, de Jonge, \& Han(2000)]{gate00}
Gatewood, G., de Jonge, J. K., \& Han, I. 2000, \apj,
533, 938

\bibitem[Grevesse \& Noels(1993)]{gn93} Grevesse,
N., \& Noels, A.  1993, in Origin and Evolution of the
Elements, ed. M. Prantzos, E. Vangioni-Flam, \& M. Cass\'e
(Cambridge:  Cambridge Univ. Press), 15

\bibitem[Gratton, Carretta, \& Castelli(1996)]{gr96}
Gratton, R. G., Carretta, E., \& Castelli, F. 1996, \aap,
314, 191

\bibitem[Grocholski \& Sarajedini(2003)]{gs03} Grocholski,
A. J., \& Sarajedini, A. 2003, preprint (astro-ph/0307503)

\bibitem[Johnson \& Knuckles(1955)]{jk55} Johnson, H. L.,
\& Knuckles, C. F. 1955, \apj, 122, 209

\bibitem[Johnson, MacArthur, \& Mitchell(1968)]{jmm68}
Johnson, H. L., MacArthur J. W., \& Mitchell R. I. 1968,
\apj, 152, 465

\bibitem[Johnson et al.(1966)]{jmi66} Johnson, H. L.,
Mitchell R. I., Iriarte B., \& Wisniewski W. Z. 1966,
Comm. Lun. Plan. Lab. 4, 99

\bibitem[van Leeuwen(1999)]{vl99} van Leeuwen, F. 1999,
\aap, 341, 71

\bibitem[van Leeuwen \& Evans(1998)]{le98} van Leeuwen,
F., \& Evans, D. W. 1998, \aaps, 130, 157

\bibitem[van Leeuwen, Alphenaar, \& Meys(1987)]{vam87}
van Leeuwen, F., Alphenaar, P., \& Meys, J. J. M. 1987,
\aap, 67, 483

\bibitem[Lejeune et al.(1998)]{lejeune98} Lejeune, T.,
Cuisinier, F., \& Buser, R. 1998, \aaps, 130, 65

\bibitem[Lejeune \& Schaerer(2001)]{ls01} Lejeune, T., \&
Schaerer, D. 2001, \aap, 366, 538

\bibitem[Makarov(2002)]{makarov02} Makarov, V. V. 2002,
\aj, 124, 3299

\bibitem[Mendoza(1967)]{men67} Mendoza, E. E. 1967,
Ton. Tac. 4, 149

\bibitem[Mermilliod et al.(1997)]{mer97} Mermilliod, J.-C.,
Turon, C., Robichon, N, Arenou, F., \& Lebreton, Y. 1997,
in Hipparcos Venice `97, eds. B. Battrick \& M. A. C.
Perryman, (Paris: ESA), 643

\bibitem[Narayanan \& Gould(1999a)]{ng99a} Narayanan,
V. K., \& Gould, A. 1999a, \apj, 515, 256

\bibitem[Narayanan \& Gould(1999b)]{ng99b} Narayanan,
V. K., \& Gould, A. 1999b, \apj, 523, 328

\bibitem[Paulson, Sneden, \& Cochran(2003)]{paulson03}
Paulson, D. B., Sneden, C., \& Cochran, W. D. 2003, \aj,
125, 3185 (PSC)

\bibitem[Percival, Salaris, \& Kilkenny(2003)]{per03}
Percival, S. M., Salaris, M., \& Kilkenny, D. 2003, \aap,
400, 541

\bibitem[Perryman et al.(1998)]{per98} Perryman, M. A. C.,
Brown, A. G. A., Lebreton, Y., G\'omez, A., Turon, C.,
de Strobel, G. C., Mermilliod, J.-C., Robichon, N.,
Kovalevsky, J., \& Crifo, F. 1998, \aap, 331, 81

\bibitem[Persson et al.(1998)]{persson98} Persson,
S. E., Murphy, D. C., Krzeminski, W., Roth, M., \& Rieke,
M. J. 1998, \aj, 116, 2475

\bibitem[Pinsonneault et al.(1998)]{pin98} Pinsonneault,
M. H., Stauffer, J., Soderblom, D. R., King, J. R., \&
Hanson, R. B. 1998, \apj, 504, 170

\bibitem[Pinsonneault et al.(2003)]{paper1} Pinsonneault,
M. H., Terndrup, D. M., Hanson, R. B., \& Stauffer,
J. R. 2003, \apj, in press, astro-ph/0307554 (Paper I)

\bibitem[Robichon et al.(1999)]{rob99} Robichon, N.,
Arenou, F., Mermilliod, J.-C., \& Turon, C. 1999, \aap,
345, 471

\bibitem[Robichon et al.(2000)]{rob00} Robichon, N.,
Lebreton, Y., Turon, C., \& Mermilliod, J.-C. 2000,
in Stellar Clusters and Associations: Convection,
Rotation, and Dynamos, eds. R. Pallavicini, G. Micela,
and S. Sciortino, ASP Conf. Ser. 198, 141

\bibitem[Saumon, Chabrier, \& Van Horn(1995)]{scv95}
Saumon, D., Chabrier, G., \& Van Horn, H. M. 1995,
\apjs, 99, 713

\bibitem[Soderblom et al.(1998)]{sod98} Soderblom, D. R.,
King, J. R., Hanson, R. B., Jones, B. F., Fischer, D.,
Stauffer, J. R., \& Pinsonneault, M. H. 1998, \apj,
504, 192

\bibitem[Stauffer et al.(2003)]{sta03} Stauffer, J. R.,
Jones, B. F., Backman, D., Hartmann, L. W., Barrado y
Nevascu\'es, D., Pinsonneault, M. H., Terndrup, D. M., \&
Muench, A. 2003, \aj, 126, 833

\bibitem[Stello \& Nissen(2001)]{sn01} Stello, D., \&
Nissen, P. E. 2001, \aap, 374, 105

\bibitem[Taylor(1980)]{tay80} Taylor, B. J. 1980, \aj,
85, 242

\bibitem[Taylor \& Joner(1985)]{tj85} Taylor, B. J., \&
Joner, M. D. 1985, \aj, 90, 479

\bibitem[Terndrup et al.(2000)]{ter00} Terndrup, D. M.,
Stauffer, J. R., Pinsonneault, M. H., Sills, A., Yuan, Y.,
Jones, B. F., Fischer, D., \& Krishnamurthi, A. 2000, \aj,
119, 1303

\bibitem[Terndrup et al.(2002)]{terndrup02} Terndrup,
D. M., Pinsonneault, M. H., Jeffries, R. D., Ford, A.,
Stauffer, J. R., \& Sills, A. 2002, \apj, 576, 950

\bibitem[Upgren \& Weis(1977)]{uw77} Upgren, A. R., \&
Weis, E. W. 1977, \apj, 82, 978

\bibitem[Upgren, Weis, \& Hanson(1985)]{uwh85} Upgren,
A. R., Weis, E. W., \& Hanson, R. B. 1985, \aj, 90, 2039

\bibitem[Weis, Delucca, \& Upgren(1979)]{weis79} Weis,
E. W., Deluca, E. E., \& Upgren, A. R. 1979, \pasp,
91, 766

\bibitem[Weis \& Hanson(1988)]{wh88} Weis, E. W., \&
Hanson, R. B.  1988, \aj, 96, 148

\bibitem[Weis \& Upgren(1982)]{wu82} Weis, E. W. \& Upgren,
A. R. 1982, \pasp, 94, 474

\bibitem[Weiss \& Salaris(1999)]{ws99} Weiss, A., \& Salaris,
M. 1999, \aap, 346, 897

\bibitem[VandenBerg \& Clem(2003)]{vc03} VandenBerg, D. A.,
\& Clem, J. L. 2003, \aj, 126, 778

\bibitem[VandenBerg et al.(2000)]{dvb00} VandenBerg, D. A., 
Swenson, F. J., Rogers, F. J., Iglesias, C. A., \&
Alexander, D. R. 2000, \apj, 532, 430

\bibitem[Yi, Kim, \& Demarque(2003)]{ykd03} Yi, S. K., Kim, Y.-C., \& Demarque, P. 2003 \apjs, 144, 259

\end{thebibliography}
\end{document}